\documentclass[format=acmtog, review=false, screen=true]{acmart}
\usepackage{booktabs} 

\setcopyright{none}

\setcitestyle{square}

\usepackage{xcolor}
\usepackage{bbding}

\usepackage{algorithm}  
\usepackage{algpseudocode}  
\usepackage{amsmath}  
\usepackage{sidecap}
\usepackage{wrapfig}

\usepackage{float}
\floatstyle{plaintop}
\restylefloat{table}

\newcommand{\lyd}[1]{{{#1}}}

\begin{document}

\title[N-Cloth: Predicting 3D Cloth Deformation with Mesh-Based Networks]{N-Cloth: Predicting 3D Cloth Deformation with Mesh-Based Networks}

\author{Yudi Li}
\affiliation{%
  \institution{Zhejiang University}}
\email{11921059@zju.edu.cn}

\author{Min Tang}
\authornote{\url{https://min-tang.github.io/home/NCloth/}, Corresponding author.}
\affiliation{%
  \institution{Zhejiang University}}
\email{tang_m@zju.edu.cn}

\author{Yun Yang}
\affiliation{%
  \institution{Zhejiang University}}
\email{3160102543@zju.edu.cn}

\author{Zi Huang}
\affiliation{%
  \institution{Zhejiang University}}
\email{h.tommy.tang@gmail.com}

\author{Ruofeng Tong}
\affiliation{%
  \institution{Zhejiang University}}
\email{trf@zju.edu.cn}

\author{Shuangcai Yang}
\affiliation{%
  \institution{Tencent}}
\email{yscyang@tencent.com}

\author{Yao Li}
\affiliation{%
  \institution{Tencent}}
\email{leoyaoli@tencent.com}

\author{Dinesh Manocha}
\affiliation{%
  \institution{University of Maryland at College Park}}
  \email{dm@cs.umd.edu}

  
\renewcommand{\shortauthors}{Y. Li et al.}  

\begin{abstract}
We present a novel mesh-based learning approach (N-Cloth) for plausible 3D cloth deformation prediction. Our approach is  general and can handle cloth or obstacles represented by triangle meshes with arbitrary topologies. We use  graph convolution to transform the cloth and object meshes into a latent space to reduce the non-linearity in the mesh space. Our network can predict the target 3D cloth mesh deformation based on the initial state of the cloth mesh template and the target obstacle  mesh. Our approach can handle complex cloth meshes with up to $100$K triangles and scenes with various objects corresponding to SMPL humans, non-SMPL humans or rigid bodies. In practice, our approach can be used to generate plausible cloth simulation at $30-45$ fps on an NVIDIA GeForce RTX 3090 GPU. We highlight its benefits over prior learning-based methods and physically-based cloth simulators.

 \end{abstract}

\ccsdesc[300]{Computing methodologies~Machine learning}
\ccsdesc[300]{Computing methodologies~Physical simulation}

\begin{teaserfigure}
 \includegraphics[width=\linewidth]{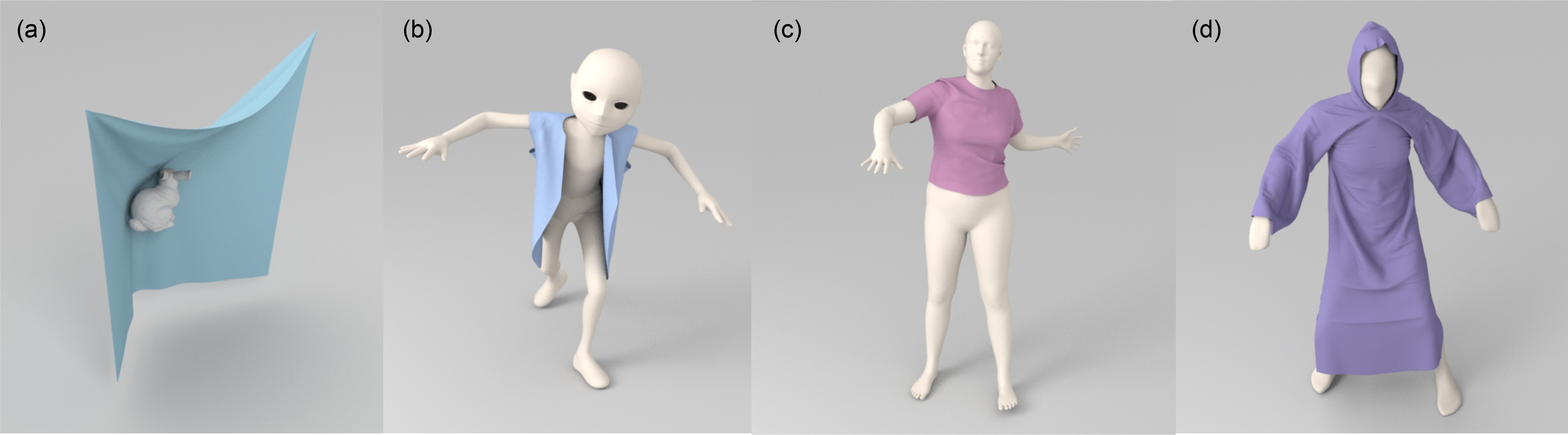}
 \centering
 \caption{{} Given the initial template of the cloth mesh and the target obstacle mesh, our network can predicate a plausible target 3D cloth mesh for general scenes. We highlight (a) the final cloth mesh wrapped around a bunny; (b) a jacket draped on a non-SMPL human body; (c) t-shirt deformation on a SMPL human body; (d) a human dressed in a robe represented by $100$K triangles. All predicted meshes are different from the datasets used for training. Our approach runs at $30-45$fps on an NVIDIA GeForce RTX 3090 GPU.
 }
\label{fig:teaser}
\end{teaserfigure}

\maketitle

\section{Introduction}

Generating plausible cloth simulation has been an active research area for many decades. The driving applications include video games and VR, computer animation, special effects, the fashion industry, virtual try-on applications, etc. There is extensive literature on simulating cloth deformation using physically-based and data-driven methods.

Physically-based methods treat cloth simulation as a deformable modeling problem and solve is using techniques from scientific computing and geometric computing. These methods  also perform collision handling for accurate simulation. The resulting algorithms can generate high-fidelity simulations and can be accelerated by exploiting GPU parallelism. However, they are mostly limited to offline simulations and are not considered fast or practical for interactive applications such as games and VR. There has been considerable interest in developing data-driven or learning-based approaches for interactive simulation.  The data-driven methods use a large number of pre-computed simulated clothing samples to synthesize cloth deformation. Recently, many learning-based methods have been proposed for draping cloth or adjusting cloth deformation to human motion~\cite{patel2020tailornet, santesteban2019learning, loper2015smpl, bertiche2020deepsd, bertiche2020physically, wu2021example, santesteban2021garmentcollisions}. While these methods can predict clothing deformation in 3D space at interactive rates, they may not work well for arbitrary scenarios with different types of objects exerting force on the cloth. In practice, these learning methods are mainly limited to predicting the deformation of clothes that conform to human movements. Moreover, many of these algorithms are limited to  SMPL-based human body  models~\cite{loper2015smpl}  or virtual try-on applications. It is not clear whether these learning methods can  extend to other types of irregular fabrics such as a table cloth  wrapping around an arbitrary obstacle like a bunny. Often these methods also require some pre-processing such as skinning ~\cite{gundogdu2019garnet,gundogdu2020garnet++}, which can introduce artifacts into subsequent network training.

\noindent {\textbf{Main Results:}}
We present a novel learning-based method (N-Cloth) to interactively predict cloth deformations in 3D. Our approach is designed for general scenes represented using triangle meshes and makes no assumption about the topology of the cloth or the shape/topology of the obstacle. Moreover, the simulation environment may consist of arbitrary rigid or deforming objects (e.g., humans in motion) that apply forces on the cloth and can result in complex deformations. Our learning method predicts the target3D cloth mesh deformation based on the initial state of the cloth mesh template and the target obstacle mesh. 

A key aspect of our learning-based approach is the use of a network that directly uses the input meshes and does not require pre-processing (e.g., mesh skinning). We extend the classic encoder-decoder architecture~\cite{4270182} with two major components: a graph-convolution-based encoder network and a fusion network. The first network transforms the input cloth and object meshes into latent vectors of a latent space and greatly reduces the input data size. This enables our algorithm to handle complex objects in the scenes defined using triangle meshes (e.g., with up to $~100K$ triangles). Our fusion network is used to derive the deforming mesh from the input cloth meshes and obstacle meshes in the latent space. This increases the accuracy of our overall learning-based method in terms of predicting arbitrary 3D cloth deformations. The connections between the outputs of an obstacle encoder and a cloth decoder are introduced to generate more accurate cloth deformations. Moreover, our learning method can also generate detailed features like wrinkles. 

We qualitatively and quantitatively analyze the performance of the proposed mesh-based network in a variety of scenarios. These include cloth meshes corresponding to many types and topologies. Furthermore, the obstacles in the scene correspond to rigid objects or a human body. In practice, our approach can generate plausible cloth deformations for all these scenarios, even when the predicted meshes are different from the datasets used in training. We also compare the performance with  TailorNet~\cite{patel2020tailornet}, a SMPL-based network, and observe lower error with respect to the ground truth. 

The novel aspects of our learning-based approach include:
\begin{itemize}
\item {\textbf{A novel mesh-based network for various scenes:}} Our approach can handle arbitrary obstacle meshes. This is in contrast to recent learning-based methods that are mainly limited to parametric human models~\cite{patel2020tailornet, santesteban2019learning, loper2015smpl, bertiche2020deepsd, bertiche2020physically, wu2021example, santesteban2021garmentcollisions}.
\item {\textbf{End-to-end neural network:}} Our method can predict cloth deformation given the initial cloth template and the target obstacle mesh. We do not perform any pre-processing (e.g., skinning computations ~\cite{gundogdu2019garnet, gundogdu2020garnet++}).
\item {\textbf{Interactive speed:}} Our approach can predict cloth deformation at $30-45$ fps on an NVIDIA GeForce RTX 3090 GPU. We observe $5-8$X and $220-300$X speedups over prior GPU-based~\cite{tang2018cloth} and CPU-based physics-based simulators~\cite{narain2012adaptive, narain2013folding}, respectively.
\item {\textbf{Plausible Results:}} We have evaluated the accuracy of our approach on a large number of complex cloth deformations and observe plausible results. Compared with TailorNet, our method can predict the cloth mesh with more wrinkles.
\item{\textbf{Lower Memory Overhead:}}  Our approach can handle cloth meshes with up to $100K$ triangles on commodity GPUs.
\end{itemize}

\section{Related Work}

In this section, we give a brief overview of prior work on cloth deformation using physically-based simulation and learning-based methods.

\subsection{Physically-based Cloth Simulation}

Physically-based algorithms use explicit Euler integration~\cite{Provot95}, implicit Euler integration~\cite{baraff1998large}, iteration optimization~\cite{liu2013fast, liu2017quasi}, or projective dynamics~\cite{Bouaziz2014} to calculate  the cloth deformation under external/internal forces. Many techniques have been proposed for  robust collision handling~\cite{Bridson02, Brochu12, Tang14,govindaraju2005quick}. Impulse-based methods and impact zones~\cite{Bridson02, Harmon08, Provot97, tang2018cloth} are used for penetration handling. Recently many techniques have been proposed to accelerate these simulations using one or more GPUs~\cite{tang2016cama,li2020p}. In practice, accurate cloth simulators can generate high--fidelity simulations, and we regard them as the ground truth for our learning approach.

\subsection{Data-driven Approaches}

Many data-driven approaches have been proposed for cloth deformation synthesis~\cite{Feng10,Wang10}. \lyd{By combining high-quality wrinkles with a coarse cloth simulation~\cite{zhang2021deep, chentanez2020cloth}, visually plausible results can be generated at interactive rates. These methods commonly need to simulate coarse deformed meshes.} Furthermore, these methods need to precompute a large dataset, and their generalizability to arbitrary scenarios tends to be limited~\cite{deAguiar10,Kim13,ZBO12}.

\begin{figure*}[h]
  \centering
  \includegraphics{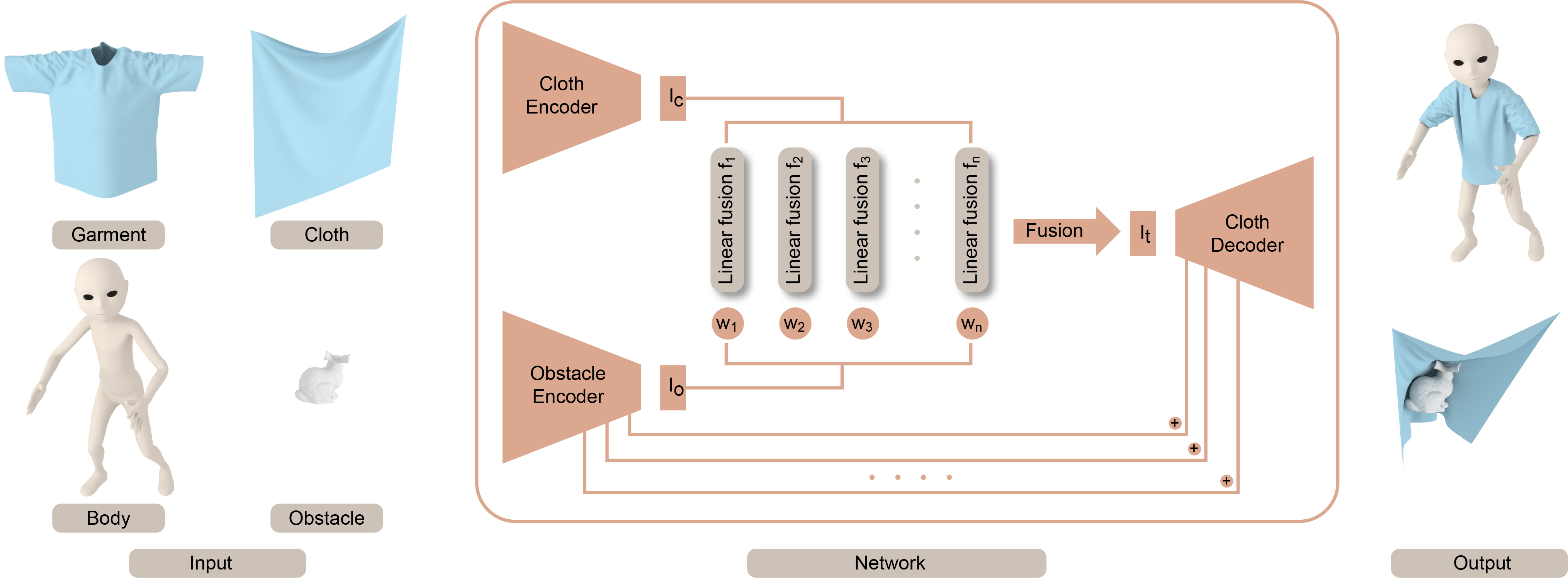}
  \caption{\label{fig:arch}{Our mesh-based network architecture:}The initial cloth mesh template is encoded into a vector $\mathbf{I}_{c}$ in the latent space by a cloth encoder. The obstacle mesh in the target state is encoded as vector $\mathbf{I}_{o}$, which is used as a weighting factor in the latent space by the obstacle encoder. With a fusion network, cloth vector $\mathbf{I}_{c}$ is weighted and fused by $n$ functions with different weights to compute the latent vector $\mathbf{I}_{t}$. $f_i, f_2, ..., f_n$ are linear fusion functions with trainable parameters. Finally, the latent vector is restored to the deformed cloth mesh in the target state by the cloth decoder. The output of each layer in the obstacle encoder is connected to the cloth decoder by a linear function to add more obstacle impact.}
\end{figure*}

\subsection{Learning-based Algorithms}

Recently, learning-based algorithms have been proposed for predicting cloth deformation in 3D. Using the synthetic training data generated  using physics-based simulators, learning-based approaches can predict cloth deformation at interactive rates on commodity GPUs. \lyd{A large number of learning-based algorithms~\cite{patel2020tailornet, santesteban2019learning, loper2015smpl, bertiche2020deepsd, bertiche2020physically, wu2021example, santesteban2021garmentcollisions,wang2019learning, corona2021smplicit} have been designed for specific or parametric obstacle models such as SMPL-based~\cite{loper2015smpl} or skeleton-based human body models.} This makes it difficult to use these methods in environments with arbitrary rigid or deformable objects that can interact with the cloth.  

\lyd{Some learning-based algorithms are not limited to SMPL models. These approaches aim to process human bodies with skeleton.} Holden et al.~\cite{holden2019subspace} obtain the vector of the vertex attributes of in the subspace through PCA and divide the deformation into linear and nonlinear to make assumptions and to predict the subsequent deformation.
The GarNet network architecture proposed by Gundogdu et al.~\cite{gundogdu2019garnet,gundogdu2020garnet++} predicts the cloth deformation from the target posture with DQS (dual quaternion skinning) pre-processing 
~\cite{Kavan07} from the initial state and uses it as the final cloth deformation. \lyd{~\cite{zhang2021dynamic} learns to generate rendered characters and cloth on posed skeleton joints. All these methods focus on human bodies with skeleton and can not handle scenes without human skeletons. Our aim is to find an approach capable of processing many scenes with human meshes and other rigid body meshes.}

\section{Our Approach}

\subsection{Overview}

Our goal is to predict the target deformed cloth mesh based on the target obstacle mesh and the initial cloth mesh. We assume that they are represented as 3D meshes.  The initial cloth mesh provides a template for deformation. The target obstacle mesh is used to guide deformation. We do not make any assumptions about the initial topology of the cloth, though it remains fixed during the simulation or deformation. Thus, our network is an end-to-end method for predicting the cloth deformation. Formally, our approach can be described as:
\begin{equation}\label{eq:1}
M^c_t = \mathcal{N}_{\theta}(M^c_i, M^o_t),
\end{equation}
where $M^c_t$ is the predicted deformed target cloth mesh and $M^o_t$ is the target obstacle mesh. $M^c_i$ is the initial, undeformed cloth mesh template which is undeformed and constant for one specific kind of cloth. During the prediction, the topologies of the cloth mesh and the obstacle mesh are invariable. $\mathcal{N}_{\theta}$ is the mesh-based network and $\theta$ represents the network parameters obtained by training.

We extend the classic encoder-decoder neural network architecture~\cite{4270182} to generate the 3D cloth deformation. An overview of our approach is shown in Fig.~\ref{fig:arch}. The deformation in mesh space is nonlinear and too complicated to be modeled. The nonlinearity of the mesh deformation is transformed to linear fusion in latent space through a cloth encoder and an obstacle encoder. Thus, we obtain the latent representation of the target cloth mesh by linear fusion. The target cloth mesh is obtained by a cloth decoder. We will describe the network in more detail.

\subsection{Encoder Network}

\begin{figure}[h]
  \centering
  \includegraphics{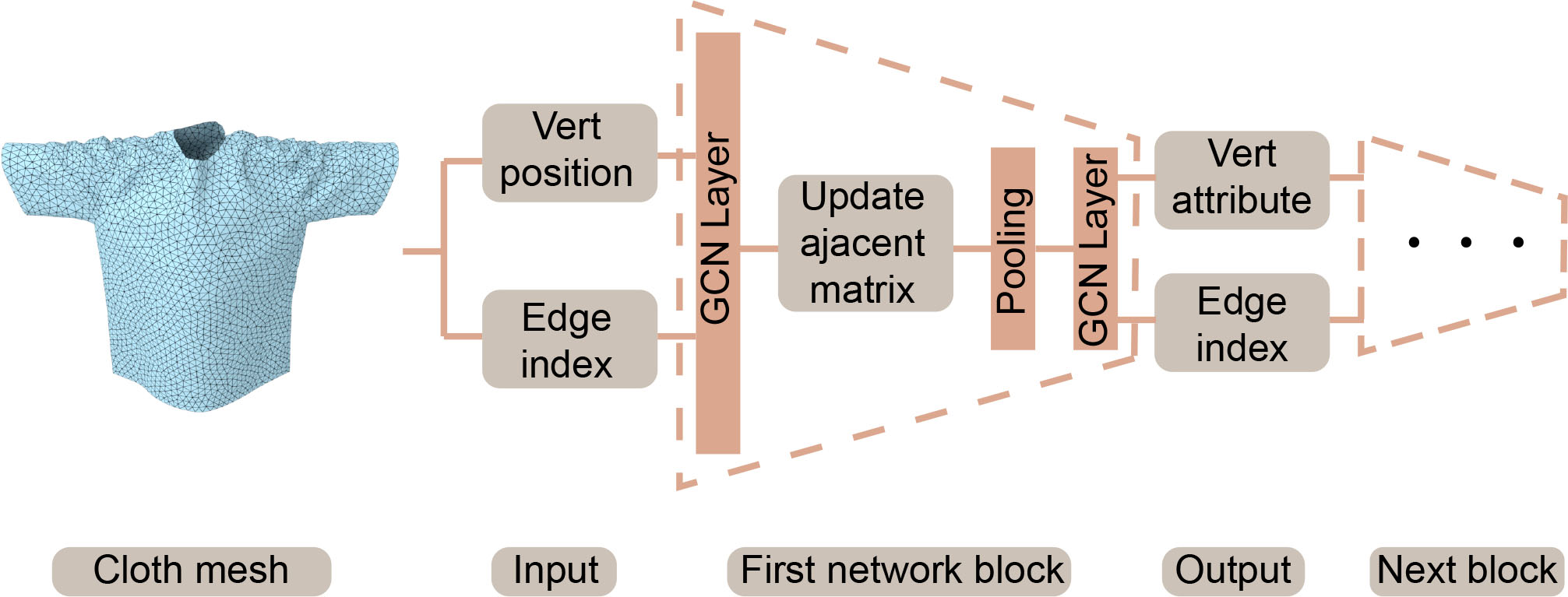}
  \caption{{The first network block of our encoder network:} Our network takes geometry information (vertex coordinates) and topology information (edge connectivity) of meshes as inputs and performs data down-sampling by outputting meshes with reduced geometry and topology information.}
  \label{fig:encoder}
\end{figure}

We use the encoder network to transform the input cloth mesh $M^c_i$ and the obstacle mesh $M^o_t$ from the 3D mesh space into a latent space. Our goal is to handle arbitrary triangle cloth meshes. A triangle mesh is similar to graph data. Therefore, the networks for processing graph data are applicable to resolving relevant mesh problems. Referring to~\cite{gao2019graph}, graph convolution is used in this paper to perform feature extraction on a mesh with abundant triangles. Both the cloth encoder and the obstacle encoder have similar network architectures.The first network block of our encoder network is shown in Fig.~\ref{fig:encoder}. Both the cloth encoder and the obstacle encoder have several network blocks similar to this initial block. Then we elaborate on the various parts of the first network block in the encoder. As shown in Fig.~\ref{fig:encoder}, we use the GCN layer~\cite{kipf2016semi} to extract features from the geometry information (vertex coordinates) and topology information (edge connectivity) of the mesh. In this manner, the GCN layer maintains the topology of the mesh. The GCN layer can be formulated as in~\cite{kipf2016semi}:
\begin{equation}
X^{(l+1)}=\sigma\left(\tilde{D}^{-\frac{1}{2}} \tilde{A}^{(l)} \tilde{D}^{-\frac{1}{2}} X^{(l)} W^{(l)}\right),
\end{equation}
where $\tilde{A}^{(l)} = A^{(l)} + I$, $A^{(l)}$ is the adjacency matrix of layer $l$. $I$ is the identity matrix for adding self-loops. $X^{(l+1)}$ and $X^{(l)}$ are the feature matrices of the layer $l+1$ and $l$, respectively. $\tilde{D_{ii}} = \sum_j \tilde{A}_{ij}$, and $W^{(l)}$ is a trainable weight matrix for layer $l$.

Since the output of the GCN layer has the same topology as the input, this formulation can result in a large number of training parameters for high-resolution meshes and thereby  exceed the GPU memory budget. To reduce the model parameters and complex non-linearity in 3D space, we use top-k-pooling~\cite{gao2019graph} to perform data down-sampling by outputting meshes with reduced topology information, i.e., with fewer vertices and connectivity information among them.

Top-k-pooling will pick $k$ vertices from the original vertex set and discard other vertices to perform down-sampling. This process may result in multiple isolated point sets, which may not work well for subsequent GCN layers because these layers extract features in terms of information related to the of the vertex. Thus, we recalculate the connectivity of mesh vertices before using a top-k-pooling layer to improve triangle connectivity. Therefore, we calculate the square of the adjacency matrix $A$ as follows~\cite{gao2019graph}:
\begin{equation}
A^{(l)}_{u}=A^{(l)} A^{(l)},
\end{equation}
where $A^{(l)}_{u}$ is the new adjacency matrix for next top-k-pooling computation. The new adjacency matrix $A^{(l)}_{u}$ corresponds to introducing more vertices around a given vertex and will reduce the isolated points after pooling.

\begin{figure}[h]
  \centering
  \includegraphics{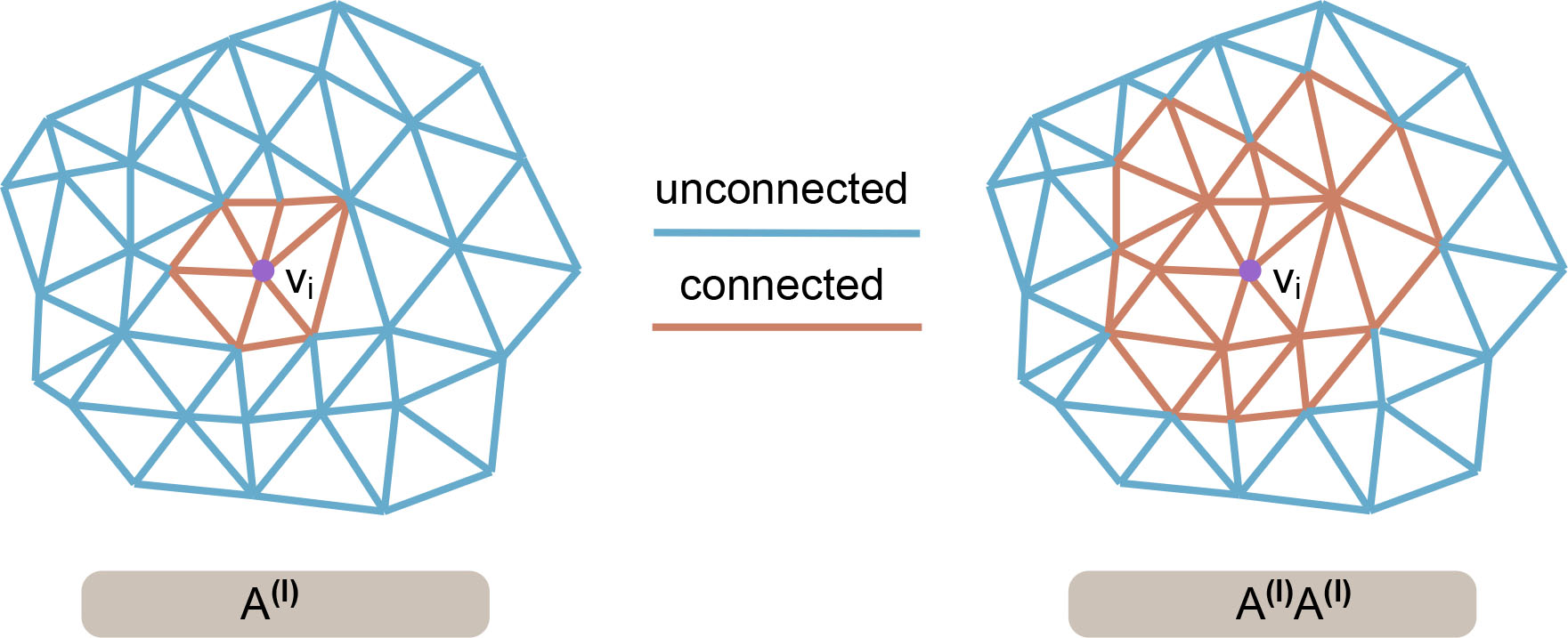}
  \caption{The connectivity of vertex $v_i$ with  adjacency matrices $A^{(l)}$ (a) and $A_{u}^{(l)}$ (b). The connectivity between vertices is strengthened with the adjacency matrix $A_{u}^{(l)}$.}
  \label{fig::al}
\end{figure}

Fig.~\ref{fig::al} shows the connectivity between a single vertex and surrounding vertices in the mesh where the adjacency matrix is $A^{(l)}$ or $A_{u}^{(l)}$. The connectivity between vertices is strengthened with the adjacency matrix $A_{u}^{(l)}$.

\subsection{Fusion Network}

The output of the cloth encoder and the obstacle encoder are $\mathbf{I}_{c}$ and $\mathbf{I}_{o}$, respectively. $\mathbf{I}_{c}$ and $\mathbf{I}_{o}$ are vectors in the latent space extracted from the cloth mesh and the obstacle mesh, respectively.  $\mathbf{I}_{c}$ and $\mathbf{I}_{o}$ are expressed as follows:
\begin{equation}
\begin{aligned}
\mathbf{I}_{c} &=\mathbf{E}_{c}(M^c_i) = (x_1, x_2, x_3, \cdots, x_m), \\
\mathbf{I}_{o} &=\mathbf{E}_{o}(M^o_t) = (w_1, w_2, w_3, \cdots, w_n), \\
\end{aligned}
\end{equation}
where $\mathbf{E}_{c}$ and $\mathbf{E}_{o}$ represent the cloth encoder network and the obstacle encoder network, respectively.
$m$ and $n$ are the dimensions of $\mathbf{I}_{c}$ and $\mathbf{I}_{o}$, respectively.
$x_1, x_2, ..., x_m$ and $w_1, w_2, ..., w_n$ are the component of latent vectors $\mathbf{I}_{c}$ and $\mathbf{I}_{o}$, respectively.

We use the fusion network to generate $\mathbf{I}_{t}$, which corresponds to the vector of the target cloth mesh in the latent space from $\mathbf{I}_{c}$ and $\mathbf{I}_{o}$.  Our formulation of the fusion network is inspired by prior work in image processing and 3D character control. Rocco et al.~\cite{rocco2017convolutional} added a correlation layer to the network for geometric matching between 2D images, and Holden et al.~\cite{holden2017phase} proposed a phase-functioned network the weights of which are updated by a phase cyclic function for 3D character control. 

We perform the fusion by linearly weighting a set of linear fusion functions $\{f_1, f_2, f_3, \cdots, f_n\}$, which all take $\mathbf{I}_{c}$ as an input. Here $\mathbf{I}_{o}$ is used as the linear weight. \lyd{The linear fusion functions are defined as follows:}
\begin{equation}
f_i(\mathbf{I}_{c}) = \begin{bmatrix}
 \alpha _{i}^{1} \\
 \alpha _{i}^{2} \\
 \cdots  \\
 \alpha _{i}^{m}
\end{bmatrix} \begin{bmatrix}
x_{1} + x_{2} + \cdots + x_{m}
\end{bmatrix},
\end{equation}
\lyd{where $\{\mathbf{\alpha}_{i}^{1}, \mathbf{\alpha}_{i}^{2},\cdots,\mathbf{\alpha}_{i}^{m}\}$ is a set of trainable coefficients} and $i \in [1,n]$. The overall fusion process can be expressed by the following formula:
\begin{equation}
\begin{aligned}
\mathbf{I}_{t}&=\sum_{i=1}^{n}{w_i f_{i}(\mathbf{I}_{c})}\\
&=w_{1} f_{1}(\mathbf{I}_{c})+w_{2} f_{2}(\mathbf{I}_{c})+w_{3} f_{3}(\mathbf{I}_{c})+\cdots+w_{n} f_{n}(\mathbf{I}_{c}).
\end{aligned}
\end{equation}
We use this formulation to obtain the vector $\mathbf{I}_{t}$ of the target cloth mesh in the latent space. Thus, we obtain a linear latent space where the deformation can be expressed as a linear fusion. In practice, we observe that our linear formulation can predict plausible results, and the errors are rather small (see  Section~\ref{sec:ret}).

In the mesh space, the influence of the obstacle mesh on the deformed cloth mesh may be complex and non-linear (e.g., due to collisions between the cloth and the obstacle). With our fusion network, we are able to model the non-linearity as weighted combinations of latent vectors and obtain the parameters of the combination function by training. In addition, the dimensions $m$ and $n$ of $\mathbf{I}_{c}$ and $\mathbf{I}_{o}$, respectively, also govern the accuracy of our predicted deformation. In our implementation, we set $m = 96$ and $n = 80$ for most benchmarks. We choose these dimensions by experiments and find that increasing them does not obviously improve the results.

\subsection{Decoder Network}

We use the decoder network to generate a cloth mesh in the world space from $I_t$. The problem of recovering the cloth mesh from the latent space has been investigated by~\cite{chen2020multi,chentanez2020cloth,chen2021deep}. Although these methods use graph convolution networks, the resulting decoder networks use the same sampling information as the encoder networks. In addition, there is almost no deformation between the input mesh and the output mesh.  However, this formulation does not work well when the target cloth mesh involves a large deformation relative to the initial mesh. In our formulation, we assume the cloth mesh maintain the same topology during deformation. As a result, we reduce the problem to only computing the geometric information of the deformed mesh (i.e., the vertex coordinates).

To compute the coordinates of the deformed vertices, we use a Multilayer Perceptron (MLP) network to decode the feature vector $\mathbf{I}_{t}$ of the target deformed cloth. In each layer of cloth decoder, we apply dropout regularization by randomly disabling 20\% of the hidden neurons to avoid overfitting the training data. The output of our decoder network is a one-dimensional vector that will be reshaped to $(num, 3)$ as vertex coordinates of the target cloth mesh, where $num$ is the number of vertices of the target cloth mesh. In addition, we add the output of each layer of the obstacle encoder to the input of the corresponding layer of the decoder through a linear layer connection. In this way, we increase the effects of obstacles on the cloth decoder and find improved results.

\subsection{Loss Functions}

Loss function is a key component of our learning-based algorithm. We use different loss terms to achieve plausible results and overcome penetrations.
We use the position information of the deformed cloth meshes as the ground truth and calculate the MSE loss between it and the network prediction output. The MSE loss on the positions can be expressed as:
\begin{equation}
\mathcal{L}_{{p}}=\frac{1}{N} \sum_{i=1}^{N} \left\|x_{p}^{i}-x_{g}^{i}\right\|_{2},
\label{eq:vertexloss}
\end{equation}
where $x_{p}^{i}$ is the position of vertex $i$ on the predicted cloth mesh, $x_{g}^{i}$ is the position of vertex $i$ on the ground mesh, $N$ is the number of vertices, and $\left\|...\right\|_{2}$ is the $L^{2}$ distance.

In addition, we also use a new type of loss to remove penetrations between the generated cloth and the obstacle. Our goal is to generate non-colliding cloth meshes. The penetration loss between the cloth mesh and the obstacle mesh can be expressed as:
\begin{equation}
\mathcal{L}_{{e}}=\frac{1}{N}
\sum_{i=1}^{N} \left(d_{\epsilon}-\min \left(\left(x_{{p}}^{i}-x_{{o }}^{i}\right) \cdot \mathbf{n}_{ {o}}^{i}, d_{\epsilon}\right)\right),
\label{eq:peloss}
\end{equation}
where \lyd{$x_{p}^{i}$ is the position of vertex $i$ on the cloth mesh} and $x_{{o }}^{i}$ is the nearest point to $x_{p}^{i}$ on the obstacle mesh. $\mathbf{n}_{{o }}^{i}$ is the normal vector of $x_{{o }}^{i}$ on the obstacle mesh. $d_{\epsilon}$ is the minimum distance of penetration. $N$ is the number of vertices of the cloth mesh. As shown in Fig.~\ref{fig:penetrate}, the penetration loss can greatly overcome penetrations between a cloth mesh and a human body. \lyd{We build the AABB tree of the obstacle to find the nearest point on the obstacle mesh. Fig.~\ref{fig:obresolution} shows the effectiveness of the penetration loss on obstacles with different numbers of triangles (0.36k, 0.64k, and 2.75k).}

To prevent self-penetrations in the generated cloth mesh, we use the following loss function:
\begin{equation}
\mathcal{L}_{{s}}=\frac{1}{N}
\sum_{i=1}^{N} \left(d_{\epsilon}-\min \left(\left(x_{ {p}}^{i}-x_{ {p}}^{j}\right) \cdot \mathbf{n}_{p}^{j}, d_{\epsilon}\right)\right), 
\label{eq:selfloss}
\end{equation}
where $x_{p}^{j}$ is the nearest vertex to $x_{p}^{i}$ on the cloth mesh. $i \ne j$ and $i, j \in [1, N]$. \lyd{We concatenate two pieces of cloth with opposite normal vectors of vertices to validate the self-penetration loss in Fig.~\ref{fig:selfpe}. However, the experiments reveal that Eq.~\ref{eq:selfloss} is limited and cannot handle all self-penetrations.}

The overall loss function used to predict the cloth deformation is:
\begin{equation}
\mathcal{L}=\mathcal{L}_{{p}} + \lambda \mathcal{L}_{{e}} + \mu \mathcal{L}_{{s}},
\label{eq:loss}
\end{equation}
where $\lambda$ and $\mu$ are blending coefficients. In practice, we use $\lambda=\mu=1$ and observe good results for all the benchmarks with these values. \lyd{Since the predictions tend to be random at the beginning of the training,  Eq.~\ref{eq:peloss} and Eq.~\ref{eq:selfloss} may result in inaccurate predictions. We use Eq.~\ref{eq:loss} to train the network at the beginning and add Eq.~\ref{eq:peloss} and Eq.~\ref{eq:selfloss} during the final training process. The number of parameters of our network and the computed gradients will hinder the scalability of training on large meshes. Therefore, Eq.~\ref{eq:vertexloss} is computed on a GPU, while Eq.~\ref{eq:peloss} and Eq.~\ref{eq:selfloss} are computed on a CPU.}

\subsection{3D Cloth Prediction}

With the trained network, we can obtain the predicted cloth mesh by inputting the initial cloth mesh and the target obstacle mesh. Since the initial  cloth mesh representation has a fixed topology for a specific cloth, we only need to input the target obstacle mesh; our network is used to predict the deformed target cloth mesh.

\section{Implementation and Performance}

\label{sec:ret}
In this section, we describe our implementation and highlight the results on many complex benchmarks. We also compare the performance with prior physics-based simulators and learning-based methods.

\subsection{Implementation}

We have implemented our algorithm on a standard PC (Ubuntu 18.04.4 LTS/Intel I7 CPU@4.2G Hz/8G RAM, NVIDIA GeForce RTX 3090 GPU). We perform both network training and cloth prediction on the same platform. Our implementation uses PyTorch 1.7.0 and Python 3.8.8 as the underlying development environment.

\begin{figure}[h]
  \centering
  \includegraphics[width=\linewidth]{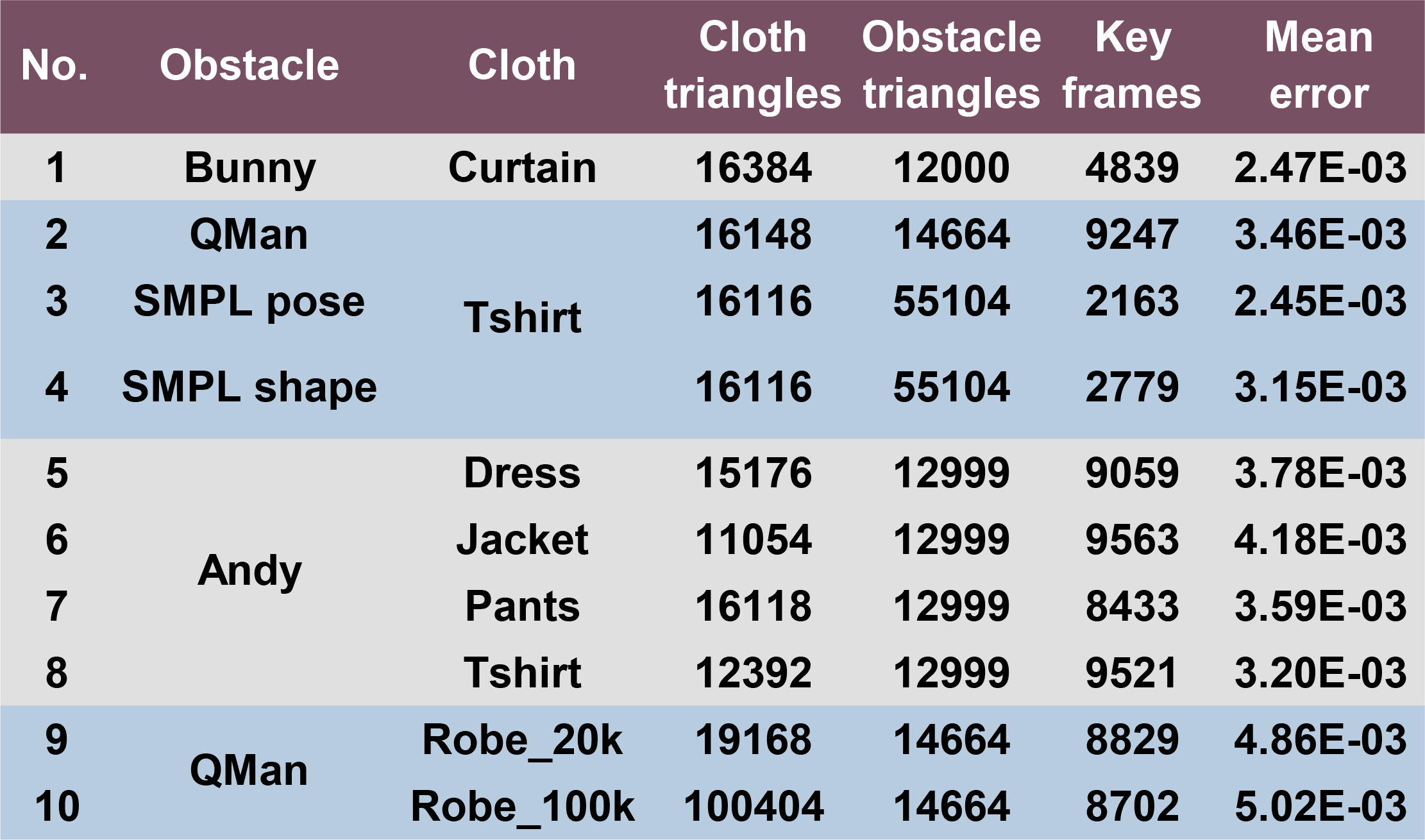}
  \caption{\label{fig:bench}We evaluate the performance of our mesh-based network on benchmarks with various characteristics: different types of obstacles, cloths with different shapes, different resolutions, etc. We highlight the number of triangles for the cloth and the obstacles,  the number of key-frames, and the mean vertex position error(in meters) for all unseen test frames between the predictions and ground truth simulated by a physics-based simulation ArcSim~\cite{narain2012adaptive}}.
 \end{figure}

{\noindent \textbf{Datasets:}}
Our mesh-based network can handle various types of cloth and obstacles. We evaluate its performance on many different cloth meshes and obstacle meshes. We consider three types of benchmark scenarios to evaluate our approach:
\begin{itemize}
\item \textbf{Rigid obstacles:} We lead a rigid bunny model through a hanging cloth from different positions in the scene. The deformation of the cloth is obtained by the simulator ArcSim~\cite{narain2012adaptive, narain2013folding, pfaff2014adaptive}. At each position, we relax the cloth for a period of time to generate a quasi-static deformation. We also simulate the results of the cloth on the bunny under different rotations and scales. The size of the bunny is scaled from 0.5 to 1.7. To ensure the uniqueness of the cloth covering on the bunny, we mark each side of the cloth as 1 or -1. We also label the side that is in contact with the bunny according to the vertex attributes of the bunny.
\item \textbf{SMPL human body model:} For the SMPL humans, we choose the representative data provided by TailorNet~\cite{patel2020tailornet}. Considering that the SMPL bodies in Tailornet have various shapes and postures, we select two types of data. One is the SMPL body data with 2779 different postures in a fixed body shape, The other is the SMPL body data with 9 different body shapes in several fixed postures. This selection of data also facilitates comparison of results from our network and Tailornet. For these human bodies, we generate their triangle meshes from the SMPL parameters.
\item \textbf{Non-SMPL human body model:} We also generate non-SMPL humans, including a child Andy and an adult male Qman. We upload the humans with canonical poses to the mixamo website\footnote{https://www.mixamo.com/}  and download about 90 different action sequences. For these action sequences, we use the physics-based simulator ArcSim~\cite{narain2012adaptive, narain2013folding, pfaff2014adaptive} to generate clothes on them. To eliminate dynamic effects, we perform linear interpolation between the adjacent poses and relax the cloth on each pose for a period of time to ensure the cloth is as static as possible. We transform all the human body meshes to the origin of the coordinates to eliminate the absoluteness of the position. The relative coordinates will enhance the generalizability of the network.
\end{itemize}

Our network can also predict different types of clothes. The child, Andy, wears different types of clothes, including a t-shirt, pants, a jacket and a dress. The jacket and dress are loose, and their deformation is different from the t-shirt and pants. These clothing simulations are also obtained by ArcSim~\cite{narain2012adaptive, narain2013folding, pfaff2014adaptive}, as above.

We evaluate the accuracy of our predicted meshes by measuring the mean error of each benchmark (as shown in Fig.~\ref{fig:bench}) with the following equation:
\begin{equation}
\mathcal{E} = \frac{ \sum_{j=1}^{M} \frac{ \sum_{i=1}^{N} \left\|x_{P}^{i, j}-x_{G}^{i, j}\right\|}{N}}{M} ,
\end{equation}
where $M$ is the number of animation key frames. $N$ is the number of vertices in the 3D mesh, and $x_{P}^{i, j}$ and $x_{G}^{i, j}$ are the positions of vertex $i$ on frame $j$. The unit of our mean error is meters. The details for each scene are shown in Fig~\ref{fig:bench}.

{\noindent \textbf{Network Training:}}  
Following \cite{patel2020tailornet} and \cite{wang2019learning}, the dataset is split for training and testing. For test data, we select 800 bunny models with different positions, rotations and scales that have not been seen during training. For the SMPL humans, we use the training and testing split provided in TailorNet~\cite{patel2020tailornet}. For other action sequences obtained from mixamo website, 90 action sequences are used during training, which produce approximately 9,000 samples. To demonstrate the generalizability of our network, during the test, we download 10 other action sequences that were unseen during training from mixamo website and predict the results of these new action poses.

Moreover, the obstacle and cloth meshes in our scene have different vertices and topologies. Thus, networks used for different scenarios have different numbers of parameters. Therefore, we train an exclusive network for each scenario. The training time for each benchmark varies from $24$ hours to $7$ days.

To accelerate the convergence of the network, we perform data normalization. We normalize the input and output vertex positions to zero mean and unit variance for all frames. During training, we uniformly reduce the learning rate from $1e-3$ to $1e-5$. We use an Adam optimizer~\cite{kingma2014adam} to train the parameters of the neural network.

\begin{figure}[h]
  \centering
  \includegraphics{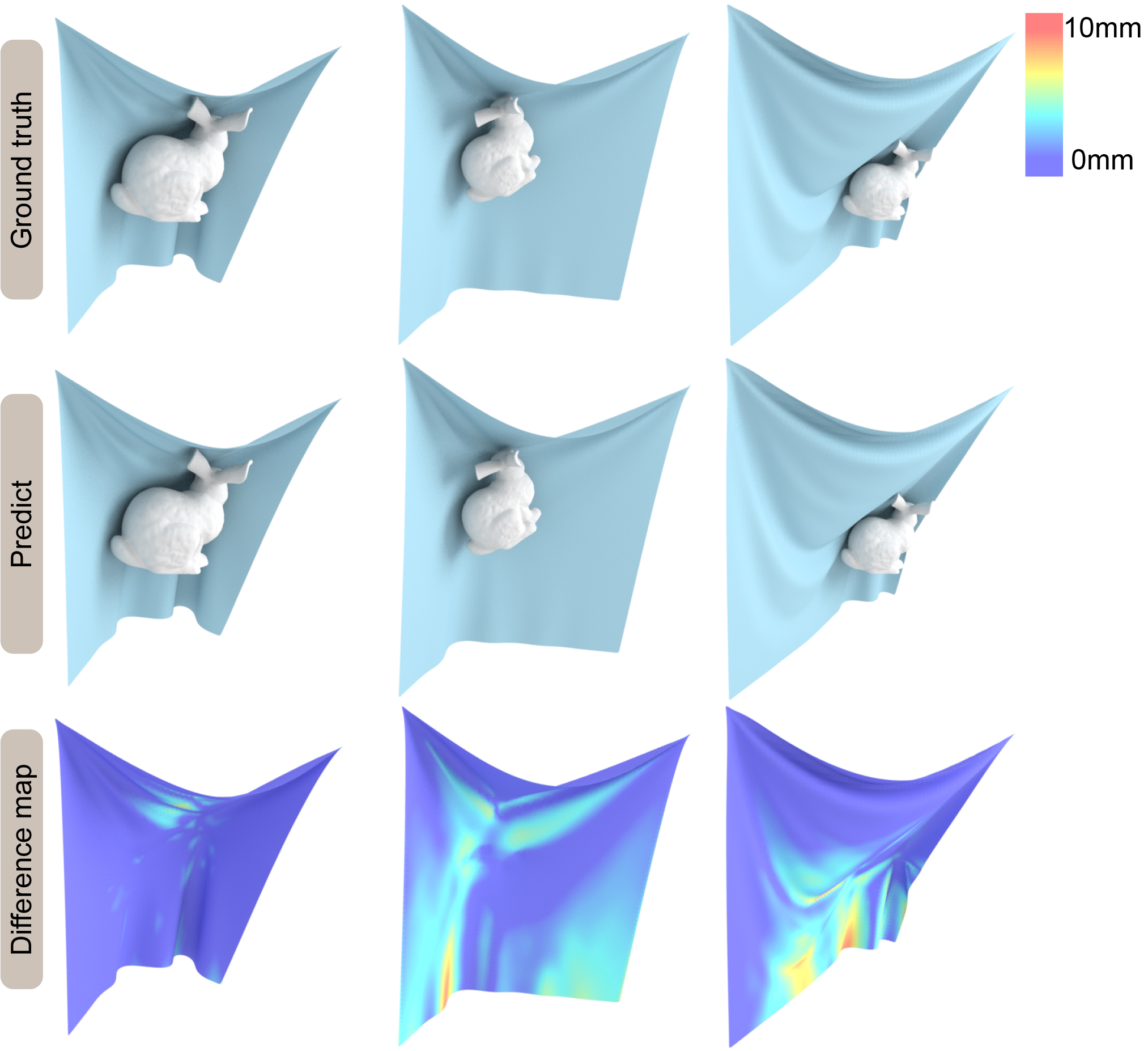}
  \caption{\label{fig:bunny}The hanging cloth draping on bunnies in different positions, rotations, and scales that are unseen in training. Compared to the ground truth computed using ArcSim (top row), our predicted meshes (second row) result in a similar mesh and visually plausible results. The deviation between the ground truth mesh and our predicted mesh is shown in the bottom row, with the error bounded by 10mm.}
\end{figure}

\begin{figure}[h]
  \centering
  \includegraphics{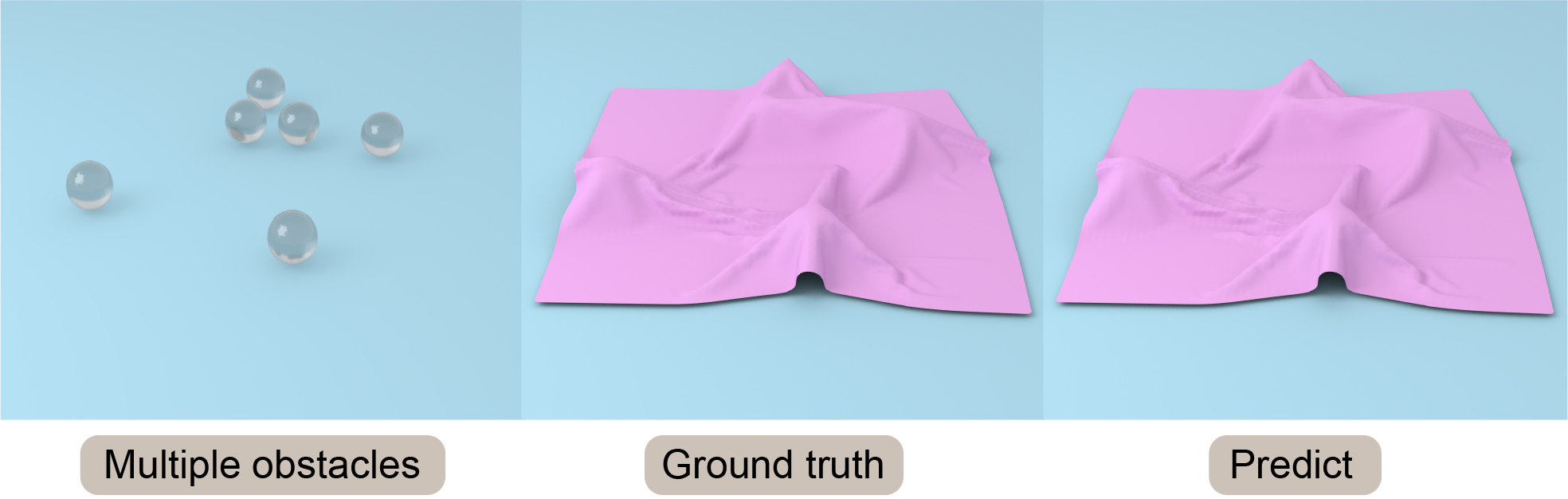}
  \caption{\label{fig:multob}\lyd{Our method works well on multiple, disjoint obstacles.}}
\end{figure}

{\noindent \textbf{Penetrations:}}
Our learning-based method uses the penetration loss function highlighted in Eq. \ref{eq:peloss}, which is designed to prevent cloth-object penetrations or cloth self-collisions.
In our benchmarks, we do not observe any deep or noticeable penetrations, though the learning-based method does not guarantee a non-penetrating final mesh. In our physics-based simulator, we use a large repulsion thickness (i.e., $1$mm) so that the training data is not only collision-free but there is some distance between non-adjacent mesh elements. This use of repulsion distance further reduces the chances of self-penetrations or collisions in the predicted cloth mesh. If the predicted mesh has a few collisions, we can solve them by simple post-processing.

\subsection{Results on Diverse Scenes}

In this section, we highlight the performance of our method on different benchmarks and compare the accuracy with physically-based simulation results. All predictions are performed on new test sets that  are different from the training data.

\begin{figure}[h]
  \centering
  \includegraphics{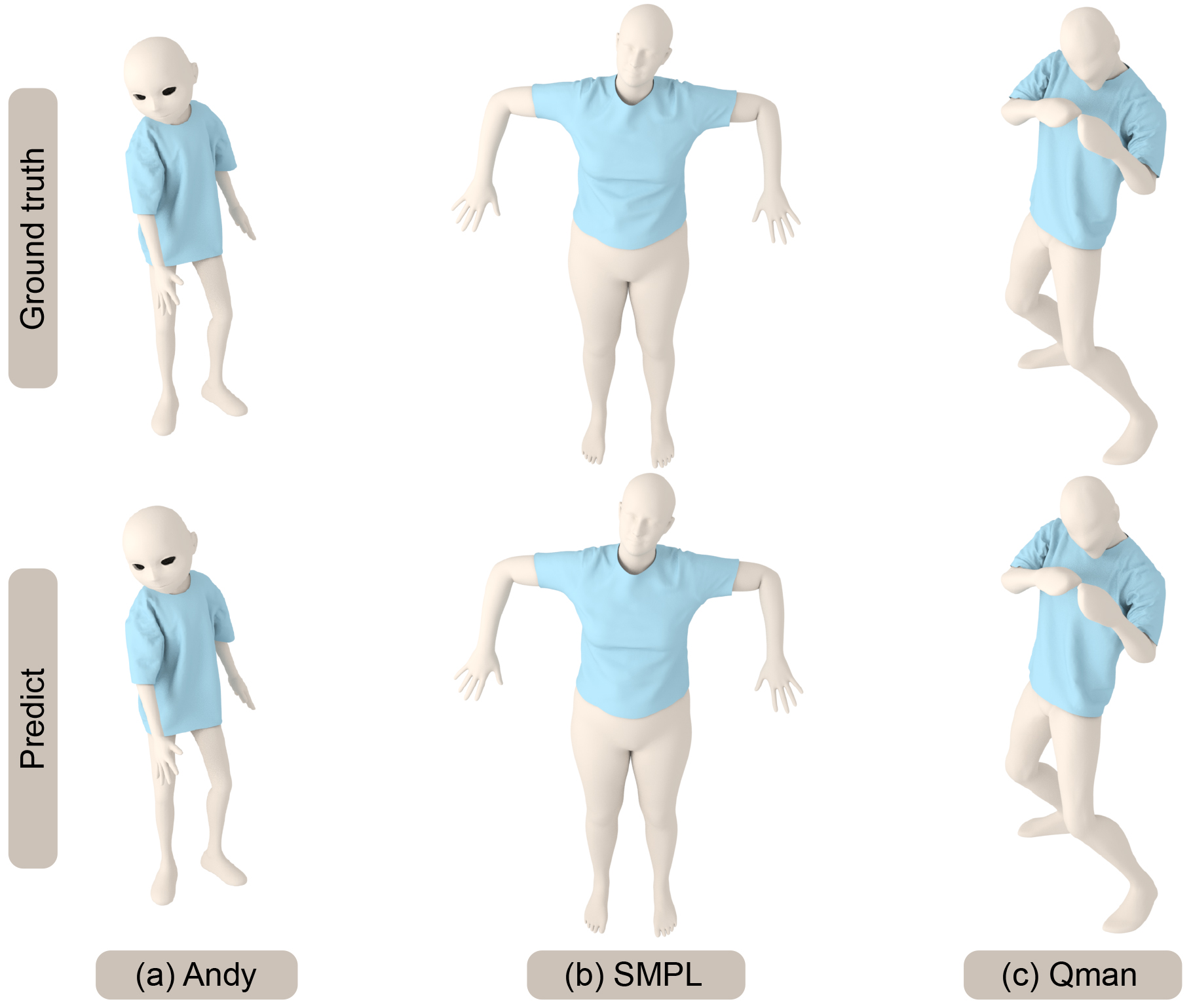}
  \caption{\label{fig:body}We show the deformation on a cloth mesh corresponding to a t-shirt on different unseen human models. All these human models are represented using triangle meshes. The human model (b) is generated from SMPL parameters. For all these benchmarks, our predictions are visually close to the ground truth.}
\end{figure}

\begin{figure}[h]
  \centering
  \includegraphics{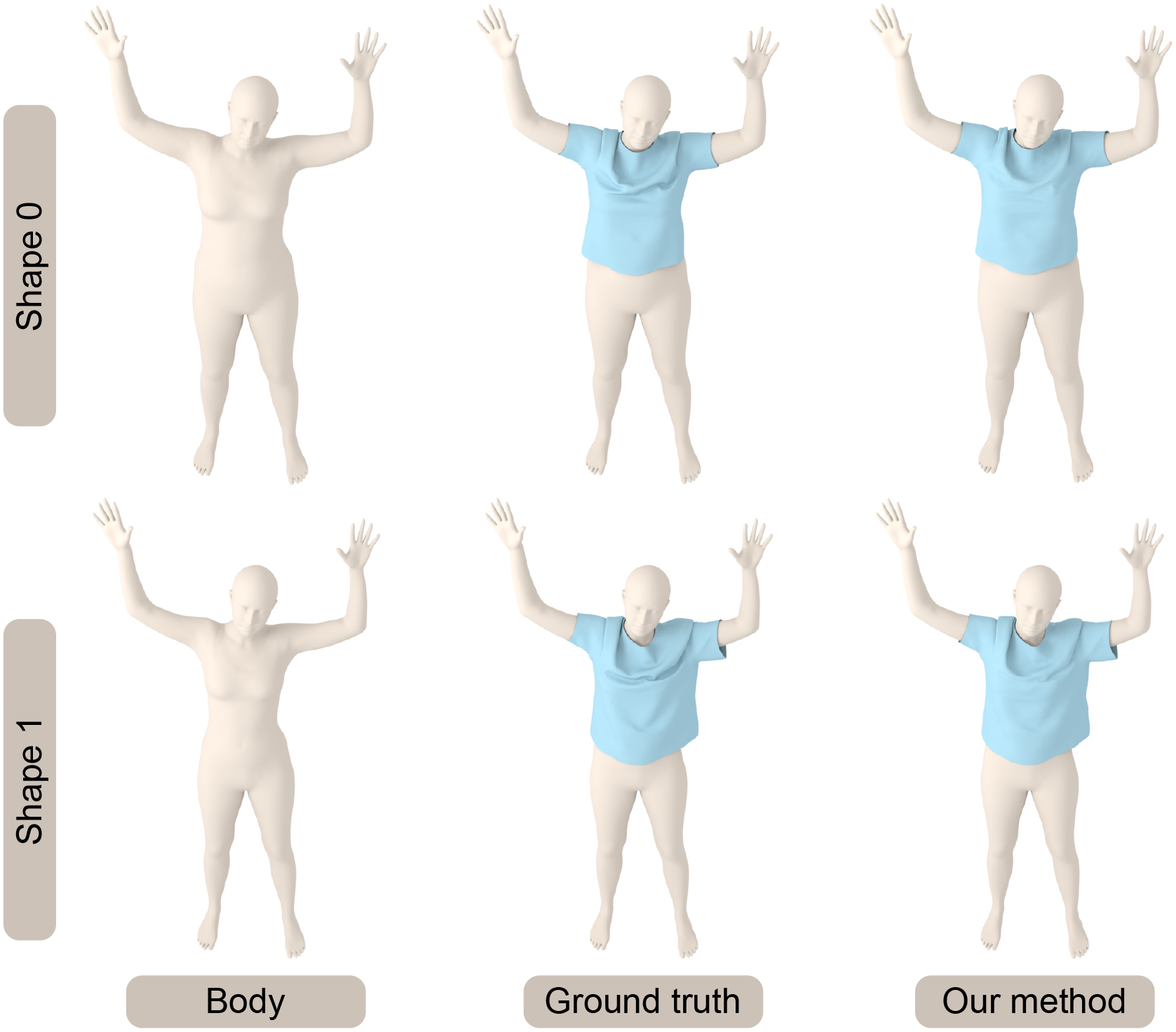}
  \caption{The ground truth and predictions of our network on body shapes unseen in TailorNet.}
  \label{fig:shape}
\end{figure}

\begin{figure}[h]
  \centering
  \includegraphics{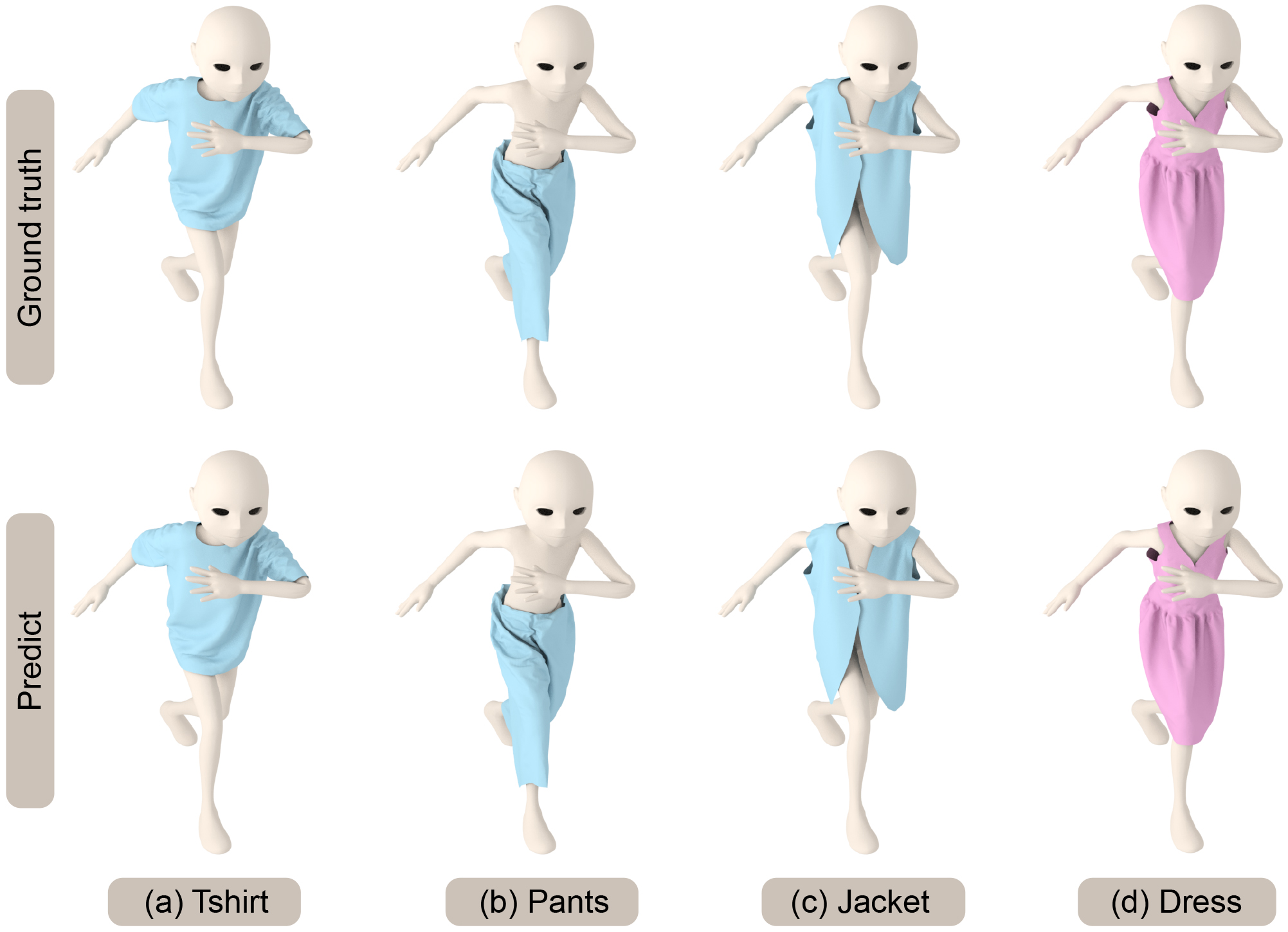}
  \caption{\label{fig:cloth}We highlight the performance of our network on different clothing types corresponding  to a t-shirt, pants, a jacket, and a dress with unseen actions. We observe that our predicted mesh is close to the ground truth and generates similar wrinkles and folds.}
\end{figure}

\begin{figure}[h]
  \centering
  \includegraphics{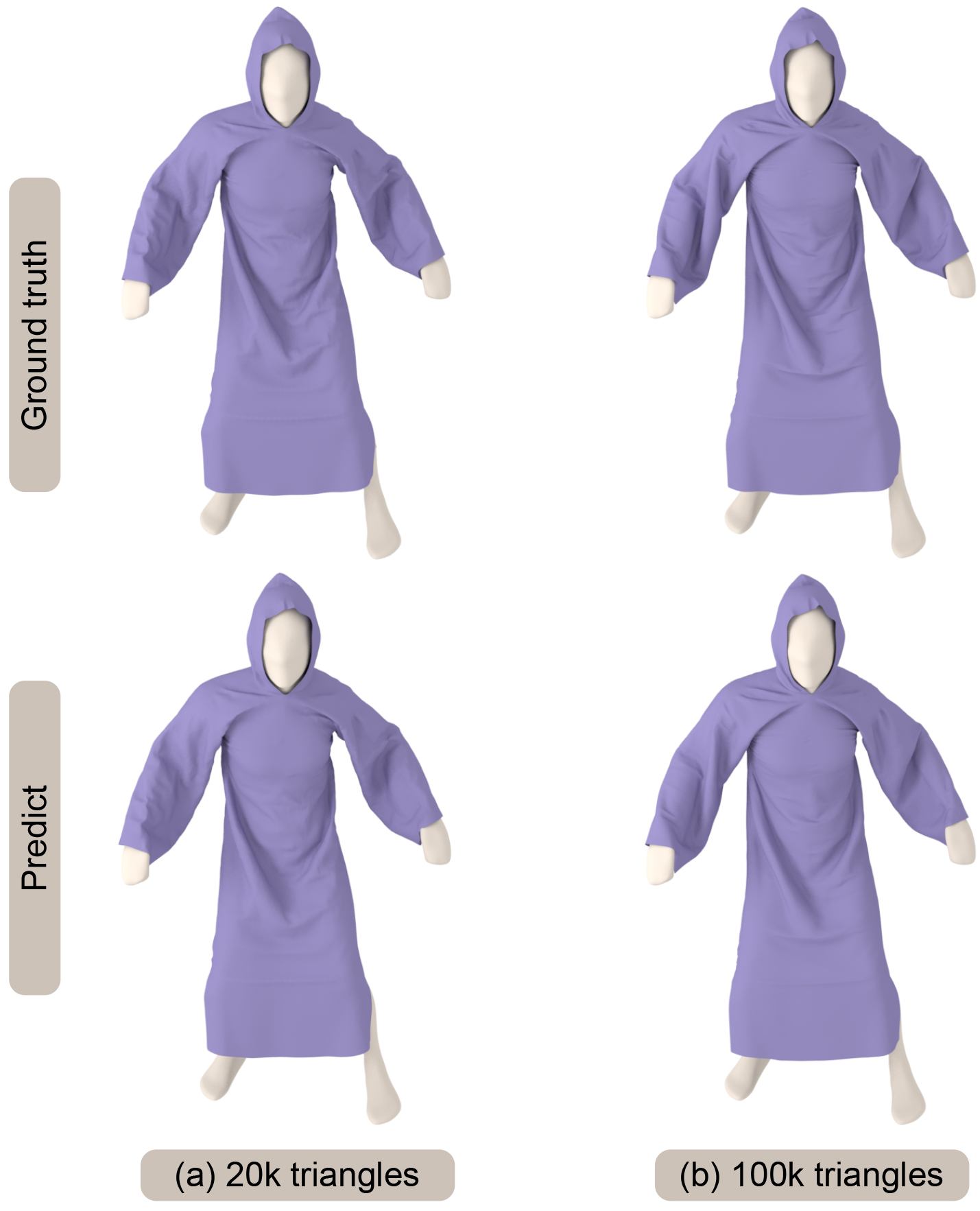}
  \caption{\label{fig:resolution}We vary the mesh resolution for a given cloth robe between $20K$ and $100K$ triangles. In each case, our predicted cloth mesh is similar to the ground truth.
}
\end{figure}

\begin{figure}[h]
  \centering
  \includegraphics{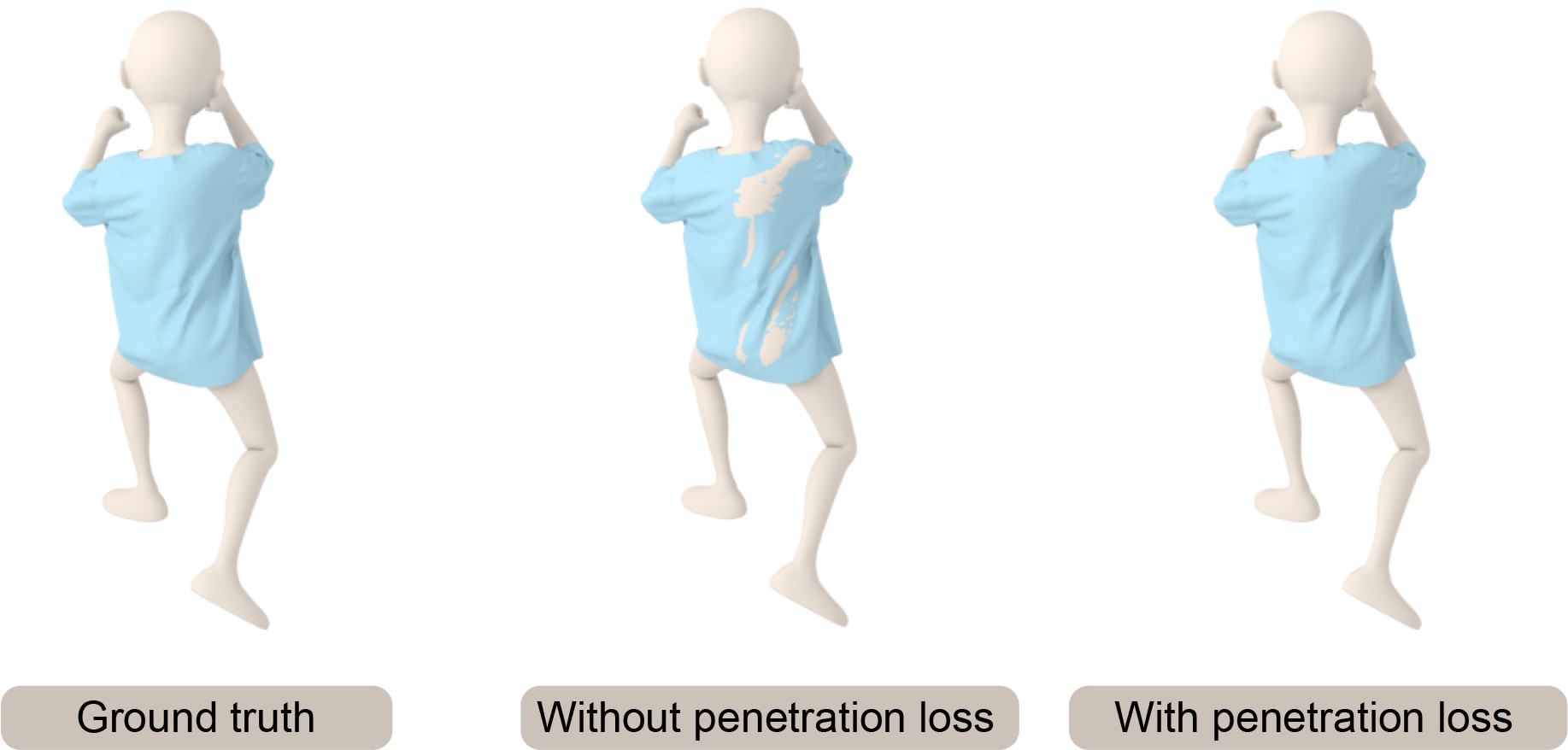}
  \caption{\label{fig:penetrate}With the penetration term in our loss function (shown in Eq.~\ref{eq:peloss}), our learning algorithm can significantly reduce the number of penetrations (shown on the right) compared to the result without penetration loss(shown in the middle).}
\end{figure}

\begin{figure}[h]
  \centering
  \includegraphics{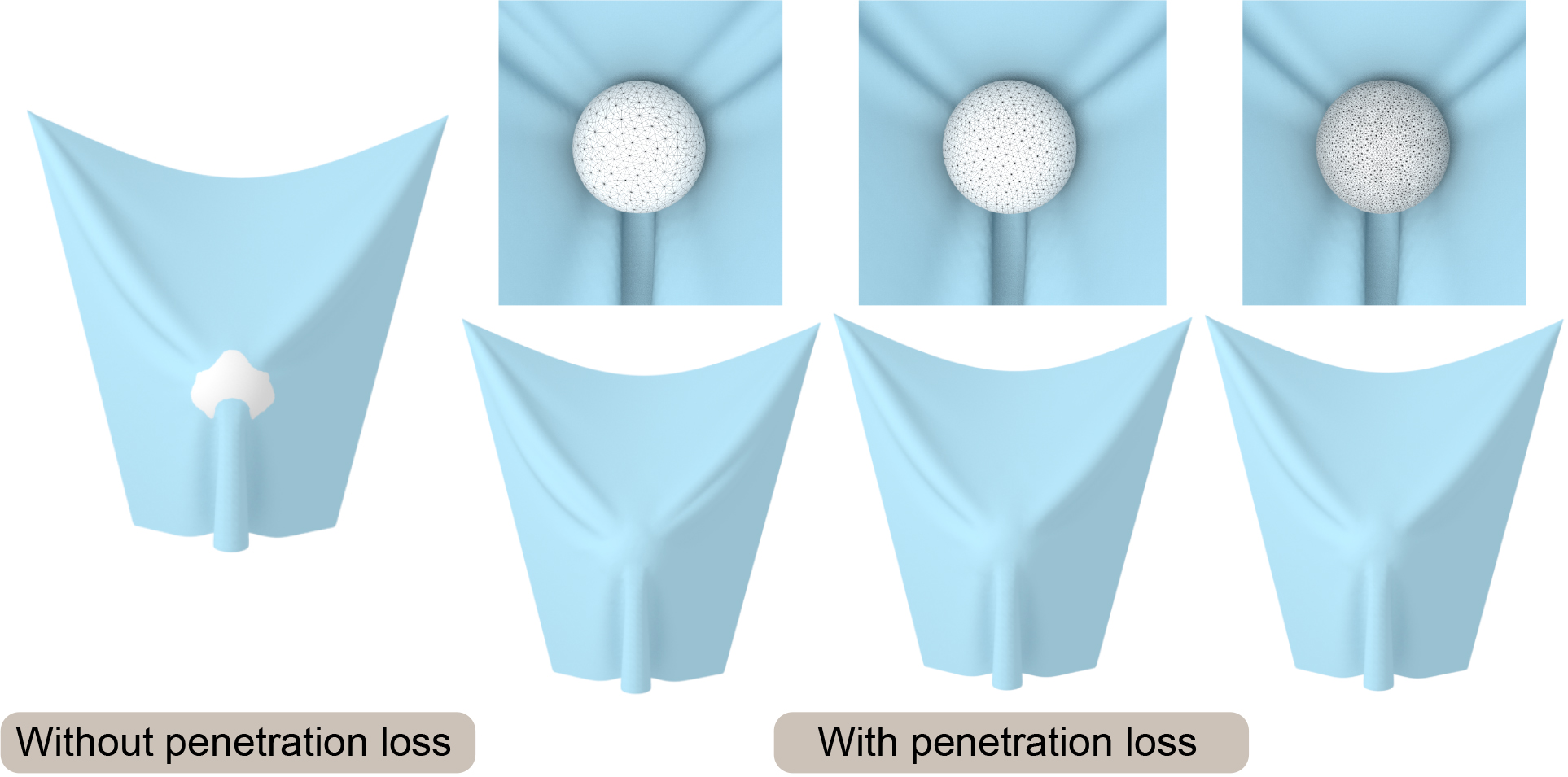}
  \caption{\label{fig:obresolution}\lyd{The results of Eq.~\ref{eq:peloss} with different obstacle discretizations.}}
\end{figure}

\begin{figure}[h]
  \centering
  \includegraphics{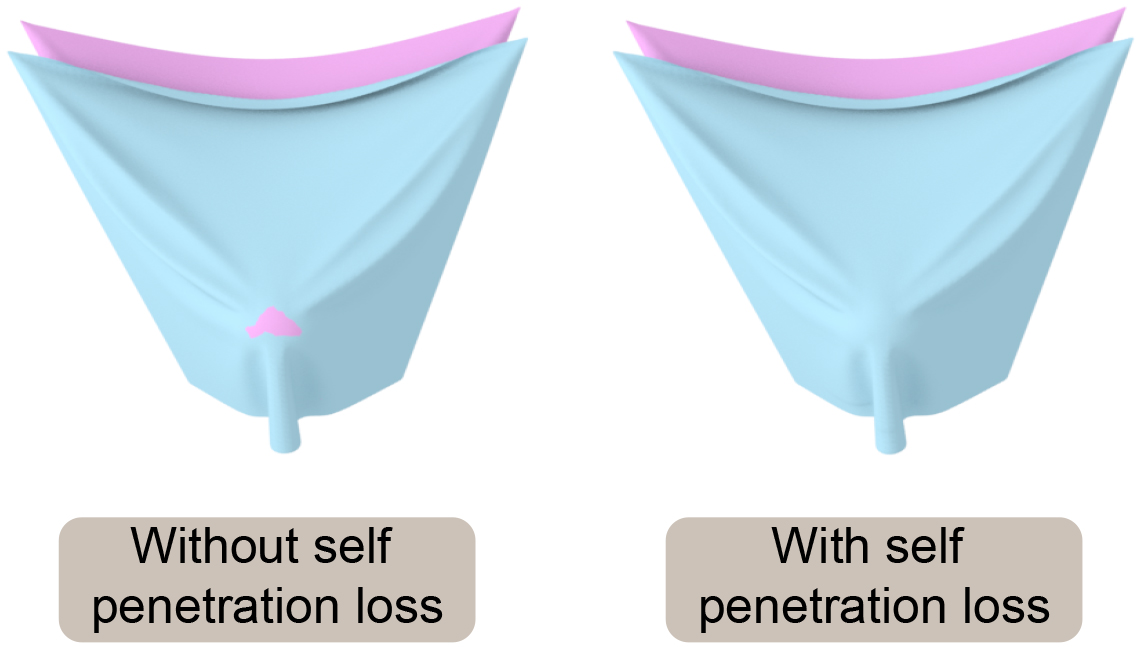}
  \caption{\label{fig:selfpe}\lyd{The results with and without the self-penetration term in the loss function (shown in Eq.~\ref{eq:selfloss}).}}
\end{figure}

Fig.~\ref{fig:bunny} highlights our results on scenes with obstacles unseen during training corresponding to  moving, rotating, or scaling rigid bodies. Our network results in favorable generalization to obstacles with unseen locations, rotations, and scales. Our approach makes no assumption about the topology of the obstacles or the cloth. We also compare the accuracy with ArcSim (an accurate physics-based simulator) and observe a high level of similarity between our predicted 3D mesh and the ground truth mesh. The mean deviation error between the vertices is less than $5$mm in our benchmarks. \lyd{If there are multiple disjoint obstacles, we combine these obstacle meshes into a single mesh and generate the cloth predictions, as shown in Fig.~\ref{fig:multob}.}
Fig.~\ref{fig:body} highlights our results on different human body models. We use the same cloth mesh corresponding to a  t-shirt on different human models. In Fig.~\ref{fig:body}, all the human bodies are represented with triangle meshes. All predictions are on an unseen test set. For example, the predictions of Andy and Qman are on the new action sequences downloaded from mixamo website. The results of SMPL are on the test data split from TailorNet. For all these benchmarks, our predicted results are visually close to the ground truth. Fig.~\ref{fig:shape} highlights the predictions of our method on unseen body shapes. \lyd{We train a single network for bodies with different shapes since these meshes have the same topologies.}
Figs.~\ref{fig:cloth} and~\ref{fig:resolution} highlight our results on cloths of different types and resolutions. For all these benchmarks, our algorithm can generate plausible results that match the ground truth meshes. All predictions from test data are totally different from the training samples.
Fig.~\ref{fig:penetrate} highlights the benefits of our penetration handling approach based on a loss function. By adding the penetration term into the loss function, our algorithm tends to alleviate the penetrations and self-collision artifacts. Although no penetration is unavailable in the predictions on the test set, it can reduce the degree of penetration and the subsequent processing work.
\lyd{Fig.~\ref{fig:obresolution} and Fig.~\ref{fig:selfpe} highlight the benefits of Eq.~\ref{eq:peloss} (with different obstacle discretizations) and Eq.~\ref{eq:selfloss}, where (self-)penetrations are effectively alleviated by these loss functions.}

To sum up, our network can not only handle SMPL and non-SMPL human bodies, but also rigid obstacles. Our network can also process various types of clothes without providing predefined skin models for those clothes. Compared with the previous method, our network can handle more scenarios and has more applications. \lyd{The predictions also show that our network can satisfactorily generalize to new, unseen data. Even when trained to predict a static deformed cloth mesh, our network generates a series of deformed cloth with fine temporal coherence on an obstacle sequence (shown in the video).}

\section{Comparisons}

In this section, we qualitatively and quantitatively compare the performance of our network with prior learning-based cloth simulation methods.

\begin{table}[h]
  \centering
  \caption{We compare the characteristics of our approach with prior learning-based methods. Some of these learning-based methods are limited to parametric human models (e.g., SMPL) and may not work for general rigid obstacles.}
  \label{tab:comp}
  \begin{tabular}{ccccccc}
    \toprule
    Method&SMPL&Non-SMPL& Triangle \\
    &body&body&mesh \\    \midrule
    TailorNet\cite{patel2020tailornet} & \color{red}\CheckmarkBold & \color{red}\XSolidBrush &\color{red}\XSolidBrush \\
    DeePSD \cite{bertiche2020deepsd} & \color{red}\CheckmarkBold & \color{red}\XSolidBrush &\color{red}\XSolidBrush \\
    \cite{santesteban2019learning} & \color{red}\CheckmarkBold & \color{red}\XSolidBrush &\color{red}\XSolidBrush  \\
    GarNet \cite{gundogdu2019garnet} & \color{red}\CheckmarkBold & \color{red}\CheckmarkBold & \color{red}\XSolidBrush  \\
    \cite{wang2019learning} & \color{red}\CheckmarkBold & \color{red}\CheckmarkBold & \color{red}\XSolidBrush \\
    \cite{bertiche2020physically} &\color{red}\CheckmarkBold & \color{red}\XSolidBrush &\color{red}\XSolidBrush \\
    DRAPE \cite{guan2012drape} & \color{red}\CheckmarkBold & \color{red}\XSolidBrush &\color{red}\XSolidBrush \\
    {Our method (N-Cloth)} & \color{red}\CheckmarkBold & \color{red}\CheckmarkBold & \color{red}\CheckmarkBold  \\
  \bottomrule
\end{tabular}
\end{table}

\subsection{Diverse Scenarios}
In Table~\ref{tab:comp}, we list the characteristics of different learning-based methods. We highlight the capabilities of different methods in terms of the kind of obstacles they can handle (e.g., SMPL models only or rigid objects).  Compared to prior methods, our approach makes no assumptions about the type or topology of the cloth or the obstacles in the scene. Most previous methods~\cite{patel2020tailornet, bertiche2020deepsd, santesteban2019learning, bertiche2020physically} are based on the SMPL model, which limits results to the SMPL human model. \lyd{Other methods~\cite{gundogdu2019garnet, guan2012drape, wang2019learning} are limited to human models represented using joints. Although they can handle the non-SMPL human body, they are unable to process other obstacles such as a bunny. ~\cite{holden2019subspace} is a complimentary method that uses PCA and subspace-only physics simulation. However, it recurrently inputs the previous prediction and accumulates errors. This makes the predicted cloth mesh appear flat with fewer wrinkles. Our mesh-based method overcomes these limitations and can handle multiple, disjoint obstacles. Our predictions have no accumulation errors and can retain fine details like wrinkles and folds.}

\subsection{Qualitative Comparisons}

\begin{figure*}[h]
  \centering
  \includegraphics{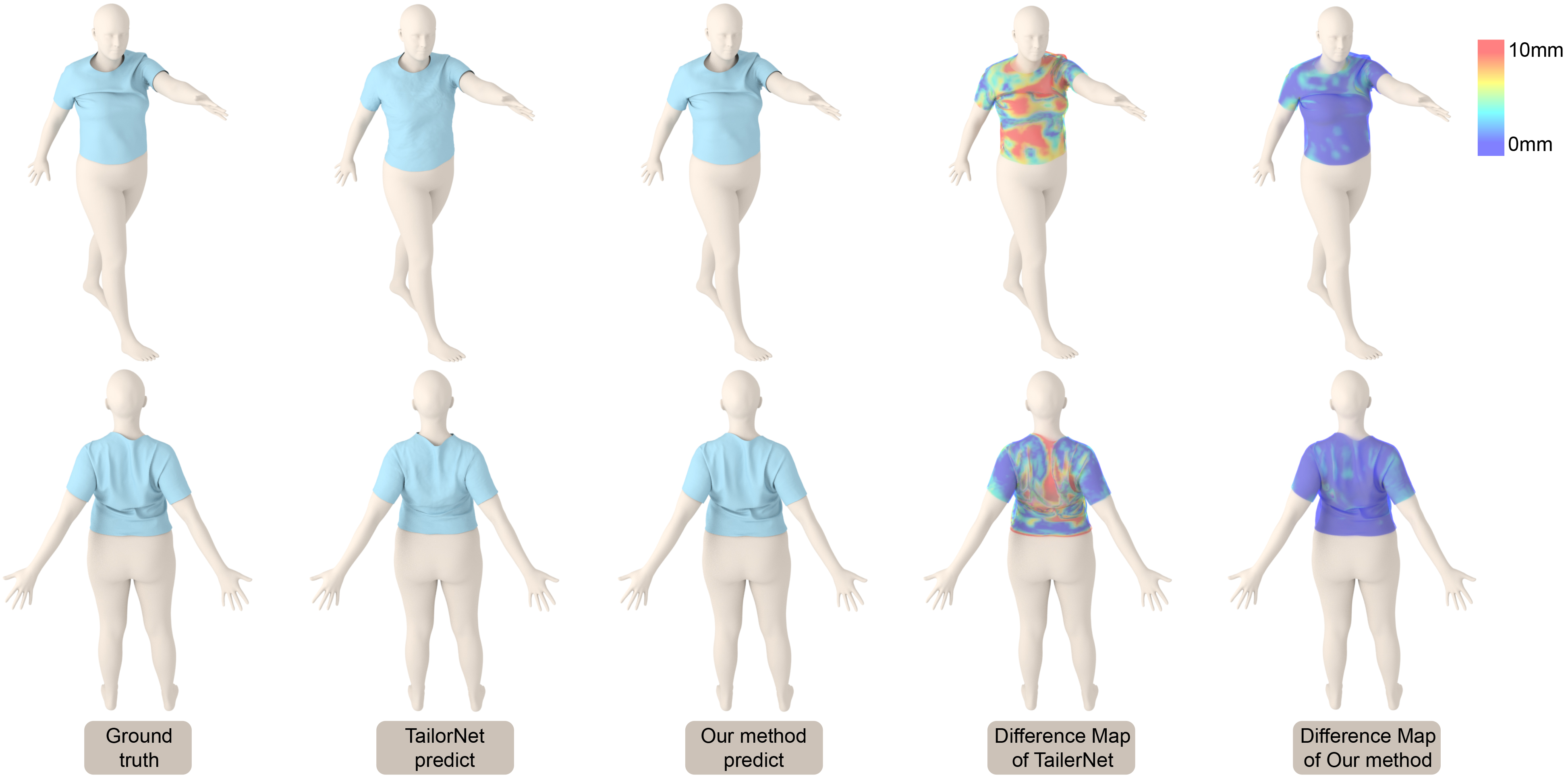}
  \caption{We compare the performance of our approach with TailorNet~\cite{patel2020tailornet} on unseen poses. We use the same dataset, available as part of TailorNet for network training. We compare the accuracy of predicted meshes generated using TailorNet and those generated using our method. We also compare the accuracy with the ground truth. We highlight the vertex error for each learning-based method by computing the difference maps with the ground truth mesh. We get results similar to TailorNet's, but with richer details.}
  \label{fig:network} 
\end{figure*}

\begin{figure}[h]
  \centering
  \includegraphics{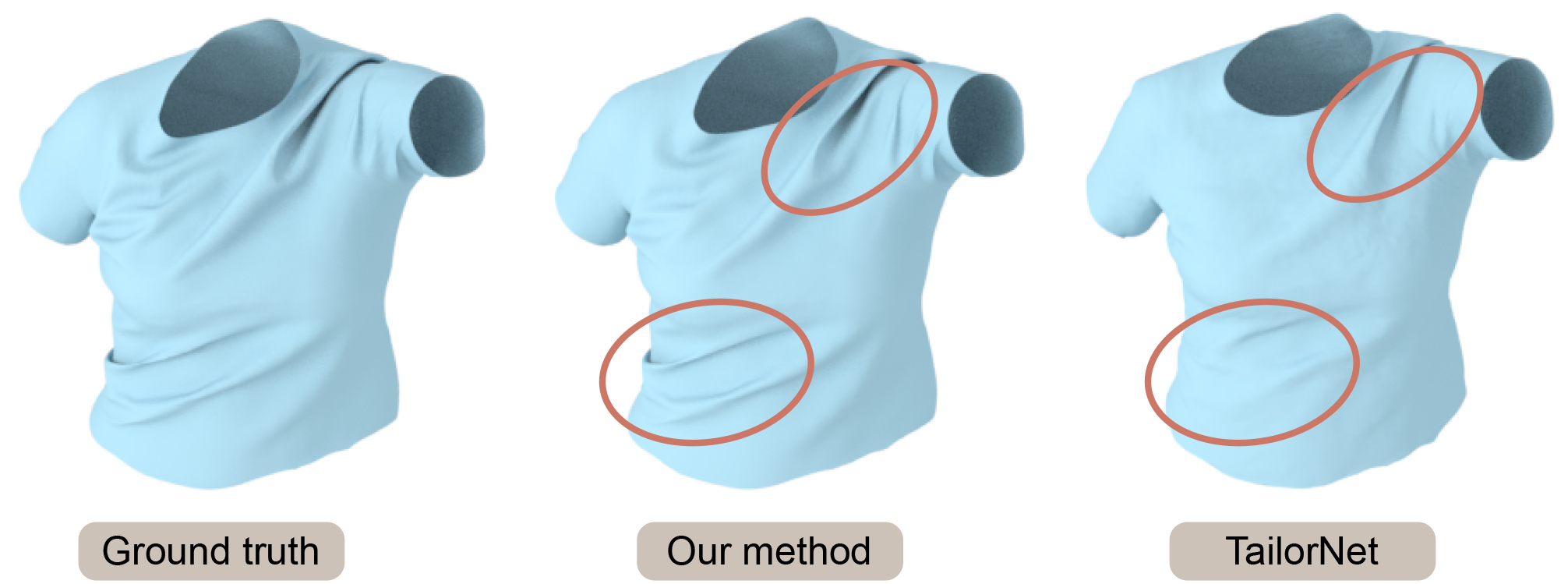}
  \caption{Our method generates better results than TailorNet in terms of preserving wrinkles, as shown in the  circled areas.}
  \label{fig:detail}
\end{figure}

We perform a detailed comparison of our method with TailorNet~\cite{patel2020tailornet}, as the code and dataset are easily available. In Fig.~\ref{fig:network}, we use the same dataset as TailorNet for network training. We compare the accuracy of the predicted cloth meshes generated by our method and TailorNet. We compute the difference maps for each mesh by comparing the results with the ground truth mesh. We observe that our method predicts similar output meshes with richer details. In addition, our method produces fewer vertex errors compared to the ground truth than TailorNet. In benchmarks with many or detailed wrinkles, we observe that our network generates better results than TailorNet, as shown in Fig.~\ref{fig:detail}. For example, our prediction of the cloth mesh has more wrinkles in the belly and shoulder areas, while TailorNet's prediction is flatter.

\begin{figure}[h]
  \centering
  \includegraphics{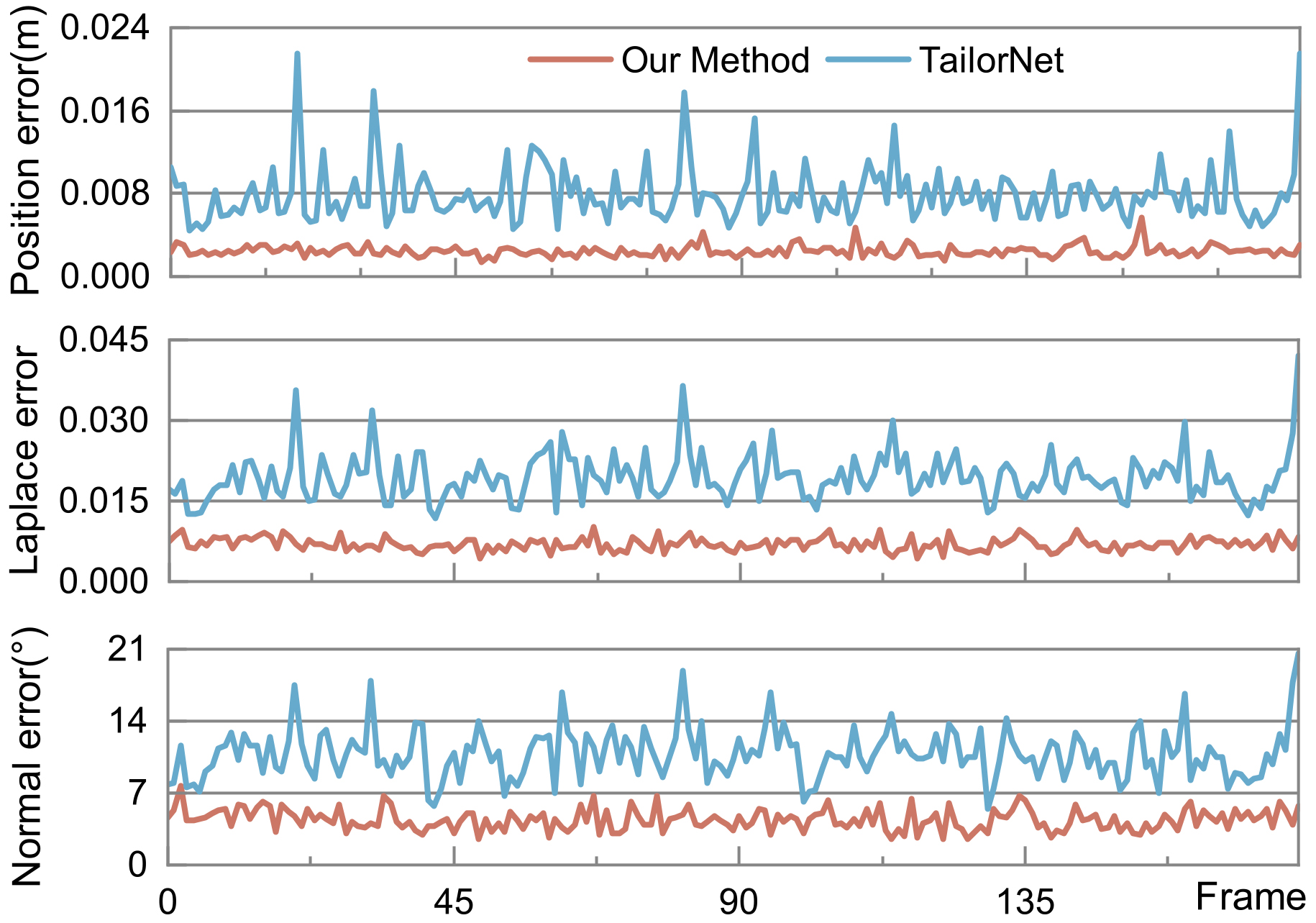}
  \caption{The errors between our network and TailorNet on the test frames.}
  \label{fig:error_figure}
\end{figure}

\subsection{Quantitative Comparisons}
We use the following error metrics for quantitative comparison between our mesh-based network and TailorNet:

\begin{equation}
\begin{aligned}
\mathcal{E}_{dist} &= \frac{1}{N} \sum_{i=1}^{N} \left\|x_{P}^{i}-x_{G}^{i}\right\| \\
\mathcal{E}_{lap} &= \frac{1}{N} \sum_{i=1}^{N} \left\|\Delta(P)-\Delta(G)\right\| \\
\mathcal{E}_{norm} &= \frac{1}{N} \sum_{i=1}^{N} \arccos \left(\frac{(y_{P}^{i})^{T}y_{G}^{i}}{\left\|y_{P}^{i}\right\|\left\|y_{G}^{i}\right\|}\right) \\
\end{aligned}
\end{equation}
where $x_{P}^{i}$ is the \lyd{position of vertex $i$ of the predicted mesh $P$.} $x_{G}^{i}$ is the position of its corresponding vertex on \lyd{the ground truth mesh $G$.}
$y_{P}^{i}$ and $y_{G}^{i}$ are the normal vector at $x_{P}^{i}$ and $x_{G}^{i}$, respectively.
$N$ is the number of vertices of the cloth mesh. $\Delta$ is the Laplace operator.

Figure~\ref{fig:error_figure} shows the error curve of our method and the predictions of Tailornet with ground truth on the test frames. The error is calculated as described above. From the curve, the trend of the errors of our network prediction is lower than TailorNet~\cite{patel2020tailornet}. We also calculate the error mean and variance for all test frames, as shown in Table~\ref{tab:error}. The statistical value of the prediction error of our method is significantly lower than that of TailorNet.

\begin{table}[h]
  \centering
  \caption{We compare the mean and standard deviations of mesh errors for our method and TailorNet on test frames based on the ground truth. Overall, our network results in fewer mesh deviation errors than TailorNet.}
  \label{tab:error}
  \begin{tabular}{ccc}
    \toprule
    Evaluation & TailorNet & Our Method \\
    \midrule
    mean $\mathcal{E}_{dist}$(m) & 7.90E-3 & 2.45E-3 \\
    std $\mathcal{E}_{dist}$(m) & 2.74E-3 & 0.54E-3 \\
    mean $\mathcal{E}_{lap}$ & 1.94E-2 & 6.97E-3 \\
    std $\mathcal{E}_{lap}$ & 4.44E-3 & 1.20E-3 \\
    mean $\mathcal{E}_{norm}$($^{\circ}$) & 10.81 & 4.40 \\
    std $\mathcal{E}_{norm}$($^{\circ}$) & 2.45 & 1.02 \\
  \bottomrule
\end{tabular}
\end{table}

\begin{figure}[h]
  \centering
  \includegraphics{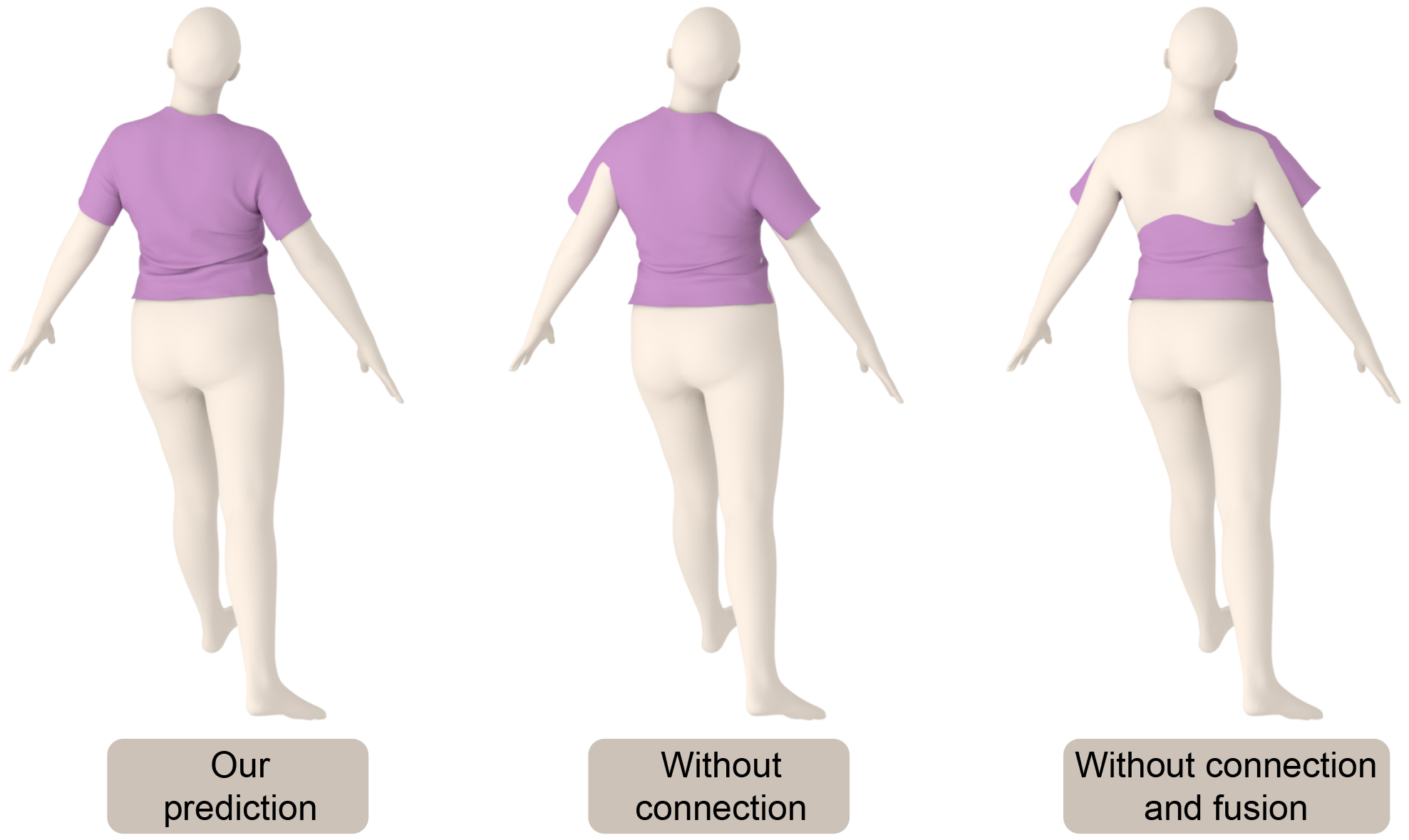}
  \caption{We discard the connections in the decoder and concatenate the cloth vector and obstacle vector in latent space as variants. The results of these variants and our network are shown above.}
  \label{fig:ablation}
\end{figure}

\begin{figure}[h]
  \centering
  \includegraphics{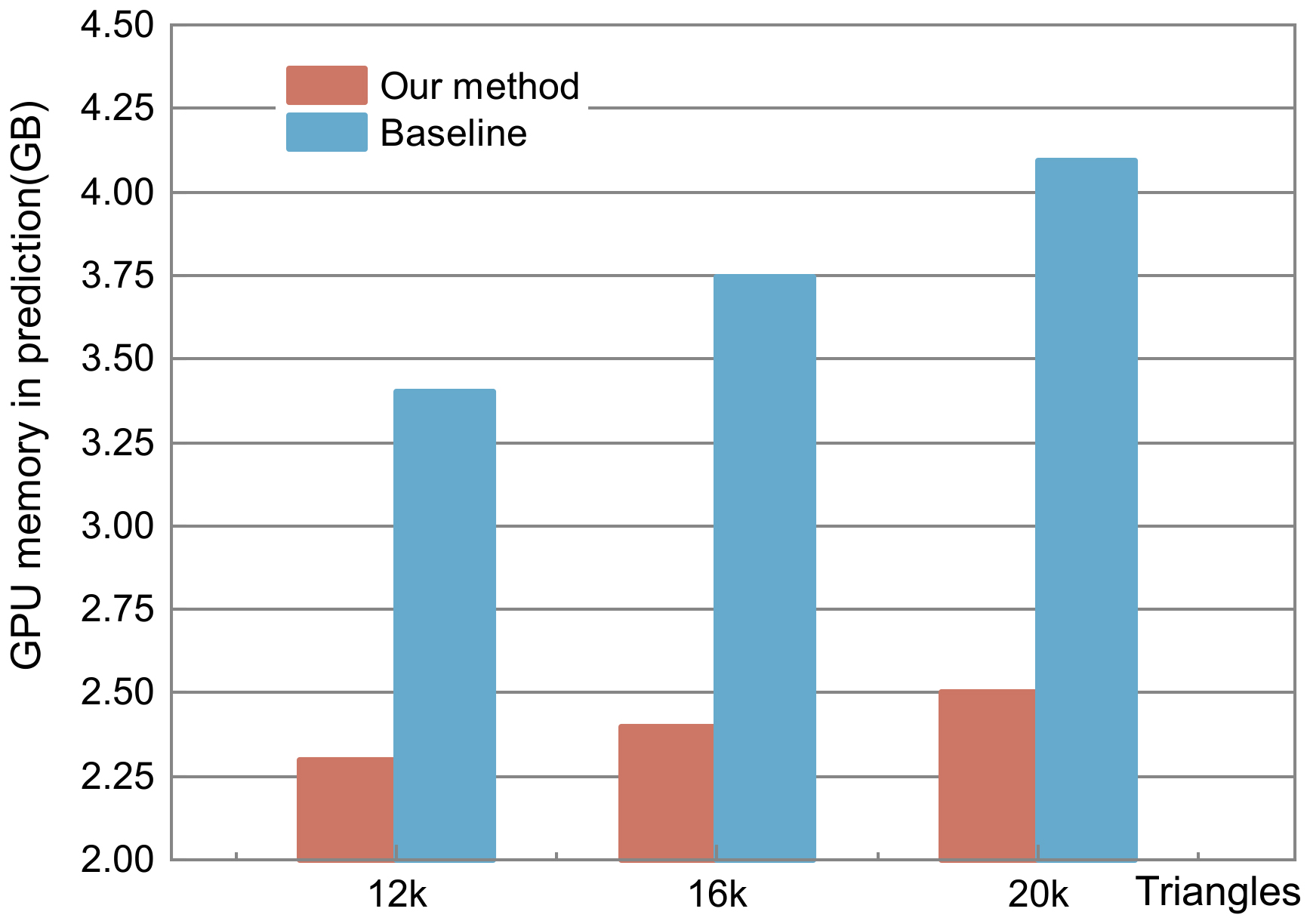}
  \caption{We implement the cloth encoder and obstacle encoder with MLP as the baseline. We compare the GPU memory of our network and the baseline while predicting a single cloth mesh at runtime.}
  \label{fig:size}
\end{figure}

\subsection{Performance Analysis}

We implement each layer of the cloth encoder and obstacle encoder with MLP as the baseline. The decoder in our network uses MLP, so we keep it invariable. Figure~\ref{fig:size} shows the GPU memory usage when our network and baseline predict a single mesh. Our network occupies less GPU memory when making predictions. In experiments, it is revealed that more GPU memory is required for baseline training, which makes it impossible to train meshes with higher resolutions. However, our method is able to train cloth meshes with more than 100k triangles.

\begin{figure}[h]
  \centering
  \includegraphics{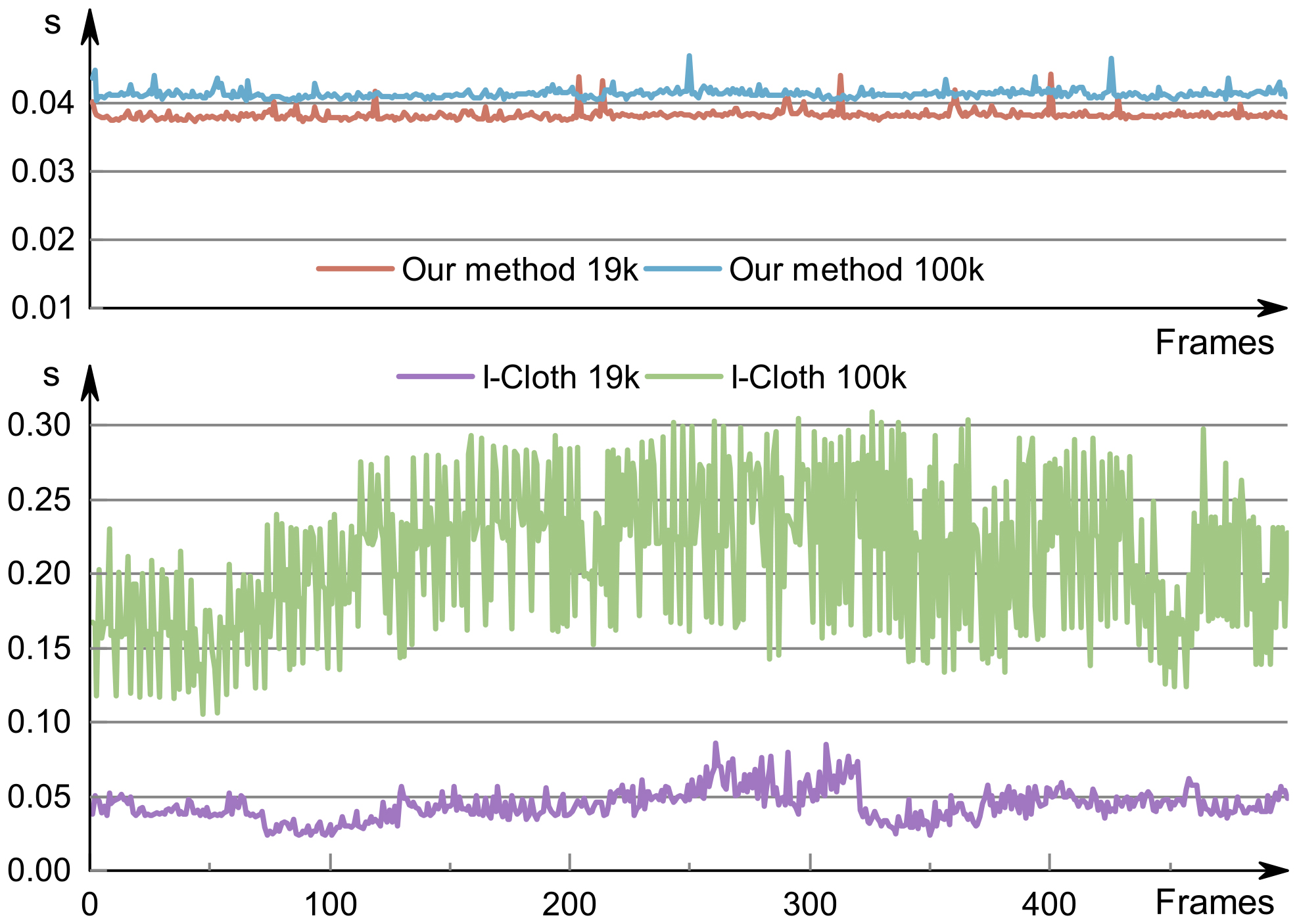}
  \caption{\lyd{Compared to} a GPU-based physics-based simulator, I-Cloth~\cite{tang2018cloth} (bottom), our method (top) results in faster performances ($5-8$X faster). We highlight the performance for cloth meshes with $19$K and $100$K triangles. The performance of our method is almost the same for a high-resolution mesh.
  }
  \label{fig:performance}
\end{figure}

We have compared the running time of our method with a GPU-based physics-based simulator called I-Cloth~\cite{tang2018cloth}, as shown in Figure~\ref{fig:performance}. Compared to I-Cloth,  our network achieves an order of magnitude performance improvement. Furthermore, the running time of our method does not change much with a higher resolution mesh. The overall accuracy and visual fidelity of the cloth mesh generated by our method are similar to those of I-Cloth.

We observe that our approach can obtain an interactive frame rate (about $30-45$ fps on  an NVIDIA GeForce RTX 3090 GPU). Compared with TailorNet, our method has no obvious advantage in running time. This is because the input of the SMPL model is a small number of parameters, while our method inputs a mesh with all the vertex and edge information. The human mesh provided by TailorNet has 55k triangles, which will increase the calculation time of our network. In practice, we concentrate more on the deformation of the cloth mesh. Therefore, we can simplify the human body mesh, which will accelerate our network.

\subsection{Ablation Experiments}

We implement a series of ablation experiments to verify the effectiveness of our network architecture. We remove the connections of each layer of the obstacle encoder in the decoder as a variant of our network. Furthermore, we discard the fusion network and use simple concatenation in the latent space as another variant. Figure~\ref{fig:ablation} shows the predictions of our network and its two variants. Our network architecture plays an indispensable role in the convergence of results.

\section{Conclusion, Limitations, and Future Work}

We present a novel mesh-based network for interactive 3D cloth prediction. Our approach is general and does not make any assumption about the topology or connectivity of the cloth or the obstacles in the scene. Our approach can handle complex cloth simulation benchmarks and predict the deformed 3D mesh at about $30-45$ fps on a commodity mesh. To the best of our knowledge, ours is the first general learning-based method that can handle arbitrary obstacle meshes and many types of cloths.

{\noindent \textbf{Limitations:}} Our approach has some limitations.
It requires considerable time to generate the training data, and it can take a few days to generate synthetic training datasets using a physics-based simulator. Furthermore, our approach assumes that the topology and connectivity of the cloth mesh is fixed. If the topology changes, we need to repeat the training step. This approach may work well for human models used for virtual try-on or dressing, as they have fixed topologies. Like prior learning-based methods, we cannot provide any rigorous guarantees in terms of absolute accuracy or collisions in our predicted mesh. \lyd{The effectiveness of our self-penetration is limited and may introduce undesirable new collisions. The computation of self-penetration depends on the discretization of the cloth mesh. We propose to different loss functions in the future to handle such self-penetrations. Or we can combine our approach with learning-based methods for collision handling~\cite{tan2021lcollision, tan2021active}.} Moreover, compared with the SMPL model which only uses a few parameters, our network performs feature extraction and fusion on the complete mesh. This results in slower performance of our network, though we observe interactive performance of 30-45fps.

There are many avenues to improve the performance in the future. Our current approaches for synthetic data generation, training, and runtime prediction are not optimized, and it is therefore possible to improve the performance. We would like to incorporate better geometric learning-based methods that can account for highly dynamic obstacles as well as small changes in mesh topology or connectivity. 
\lyd{It will be interesting to use visual knowledge~\cite{Pan21} for cloth deformation prediction.} Finally, we would like to integrate our approach with different applications corresponding to virtual try-on or gaming and evaluate the performance.





%

\bibliographystyle{ACM-Reference-Format}
\bibliography{paper}

\end{document}